\documentclass[sigconf]{acmart}
\usepackage{amsmath}
\usepackage{pifont}
\usepackage{color}
\usepackage{amsfonts}
\usepackage{subfig}
\usepackage{graphicx}
\usepackage{multirow}
\usepackage{siunitx}
\usepackage{enumerate}
\usepackage{amsfonts}
\usepackage{url}
\usepackage{hyperref}
\usepackage{algorithm}
\usepackage[noend]{algpseudocode}
\usepackage{enumerate}

\newcommand{\cmark}{\ding{51}}
\newcommand{\xmark}{\ding{55}}

\setcopyright{none} % No copyright notice required for submissions
\acmConference[Anonymous Submission to ACM CCS 2021]{ACM Conference on Computer and Communications Security}{Due 5 May 2021}{Seoul, TBD}
\acmYear{2021}

\begin{document}
\settopmatter{printacmref=false} % Removes citation information below abstract
\renewcommand\footnotetextcopyrightpermission[1]{} % removes footnote with conference information in first column

\title{Practical Privacy Attacks on Vertical Federated Learning}
\author{Haiqin Weng$^{1}$, Juntao Zhang$^{1}$,  Xingjun Ma$^{2}$, Feng Xue$^{1}$, Tao Wei$^{1}$, Shouling Ji$^{3}$,  Zhiyuan Zong$^{1}$}

\affiliation{\institution{$^{1}$Ant Group,
$^{2}$ Fudan University,
$^{3}$Zhejiang University}
\country{}}

\affiliation{\institution{\{haiqin.wenghaiqin, tilin.zjt, henry.xuef, lenx.wei, david.zzy\}@antgroup.com \\ xingjunma@fudan.edu.cn, sji@zju.edu.cn} \country{}}

%\author{
%	\IEEEauthorblockN{Haiqin Weng\IEEEauthorrefmark{1}, Juntao Zhang\IEEEauthorrefmark{1}, Feng Xue\IEEEauthorrefmark{1}, Tao Wei\IEEEauthorrefmark{1}, Shouling Ji\IEEEauthorrefmark{2},  Zhiyuan Zong\IEEEauthorrefmark{1}  \\
%	\IEEEauthorblockA{\IEEEauthorrefmark{1}Ant Group
%	\\\{haiqin.wenghaiqin, tilin.zjt, henry.xuef, lenx.wei, david.zzy\}@antgroup.com} \\
%\IEEEauthorblockA{\IEEEauthorrefmark{2}Zhejiang University \\ sji@zju.edu.cn}}
%\thanks{*work in progress}} 

% 修改的地方
% 1）Reverse Multiplication Attack攻击的假设需要自圆其说，蚂蚁应用场景举例说明；
% 2）Reverse SUM/Multiplication Attack需要增加以下Pipeline；
% 3）Reverse SUM Attack和Reverse Multiplication Attack两种攻击需要串起来。LR和XGboost实际场景尤其是纵向联邦学习场景中最常用的两种方法，我们研究了这两种方法的安全性，分别提出ReverseSUM和Reverse Multiplication  Attack。这两个攻击方法在主流联邦学习平台上都是有效的。
% 4）refer 其他的paper，包括“Privacy Preserving Vertical Federated Learning for Tree-based Models”
% 5）attack cost -- 可以delete
% 6）+ defense and related work

\begin{abstract}  
Federated learning (FL) is a privacy-preserving learning paradigm that allows multiple parities to jointly train a powerful machine learning model without sharing their private data. According to the form of collaboration, FL can be further divided into horizontal federated learning (HFL) and vertical federated learning (VFL). In HFL, participants share the same feature space and collaborate on data samples, while in VFL, participants share the same sample IDs and collaborate on features.  VFL has a broader scope of applications and is arguably more suitable for joint model training between large enterprises.  

In this paper, we focus on VFL and investigate potential privacy leakage in real-world VFL frameworks.
% We design and implement two practical privacy attacks on two real-world VFL frameworks: \emph{Reverse Multiplication Attack} (RMA) for Logistic  Regression (LR) and \emph{Reverse Sum Attack} for XGBoost. 
We design and implement two practical privacy attacks: \emph{reverse multiplication attack} for the logistic regression VFL protocol; and \emph{reverse sum attack} for the XGBoost VFL protocol. 
We empirically show that the two attacks are 
($\romannumeral1$) effective - the adversary can successfully steal the private training data, even when the intermediate outputs are encrypted to protect data privacy;
($\romannumeral2$) evasive - the attacks do not deviate from the protocol specification nor deteriorate the accuracy of the target model;
and ($\romannumeral3$) easy - the adversary needs little prior knowledge about the data distribution of the target participant. We also show the leaked information is as effective as the raw training data in training an alternative classifier. We further discuss potential countermeasures and their challenges, which we hope can lead to several promising research directions. 
%the adversary participant of a learning protocol can run reverse sum/multiplication attack against other participants and steal their private training data, even when the intermediate outputs are encrypted to protect data privacy. 
\end{abstract}  

\maketitle

\section{Introduction}
\label{sec:intro}

Federated learning (FL) has emerged as a promising privacy-aware alternative to the traditional data-sharing based machine learning. In FL, participants can jointly train a powerful machine learning model without sharing their private data~\cite{mohassel2017sp,yang2019acm,stephen2017corr,bonawitzI2018ccs}. FL has found many impactful applications ranging from the joint training of patient survival prediction models by multiple hospitals ~\cite{jochems2016distributed} to the decentralized training of powerful next word prediction models with smartphones~\cite{mcmahan2017}. 

\begin{figure*}[!th]
\centering
\subfloat[Horizontal Federated Learning]{\includegraphics[width=0.45\textwidth,height=0.2\textwidth]{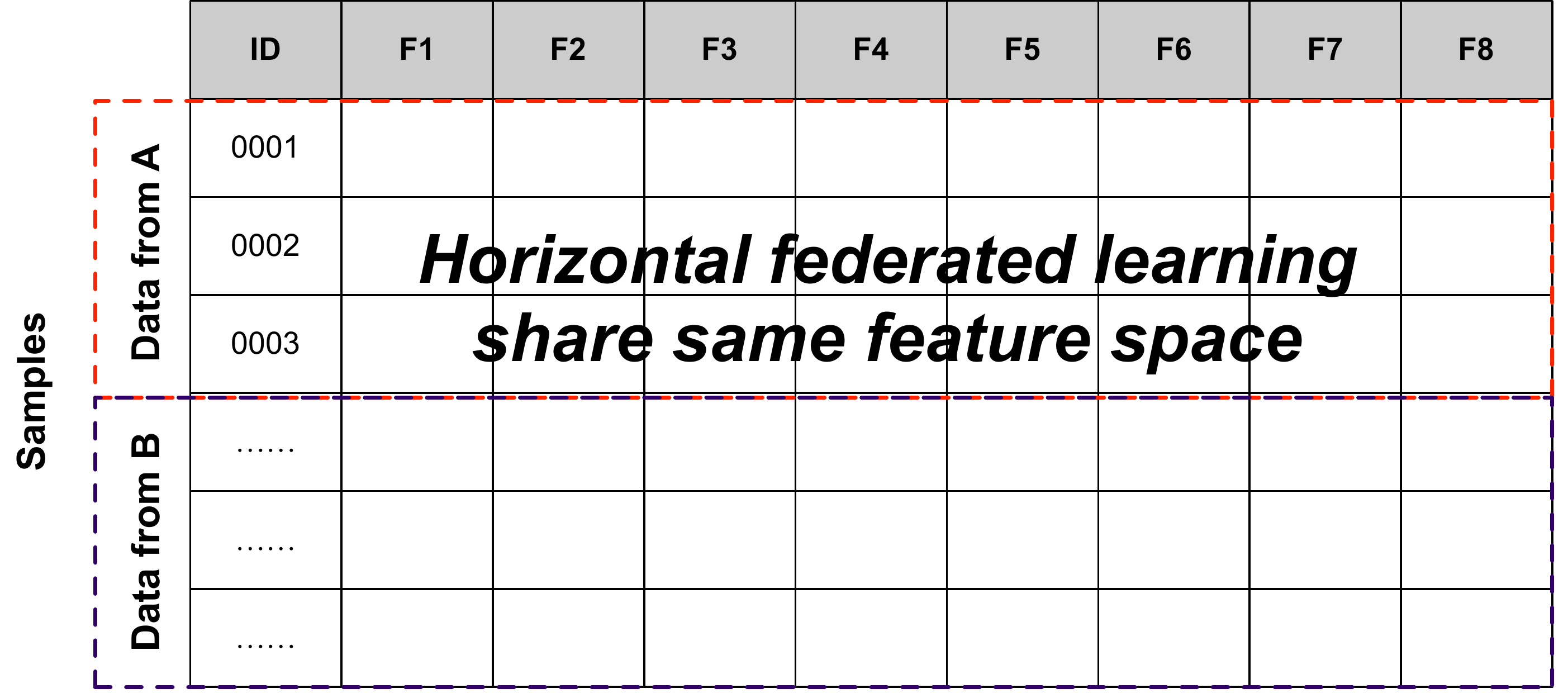}} 
\quad 
\subfloat[Vertical Federated Learning]{\includegraphics[width=0.45\textwidth,height=0.2\textwidth]{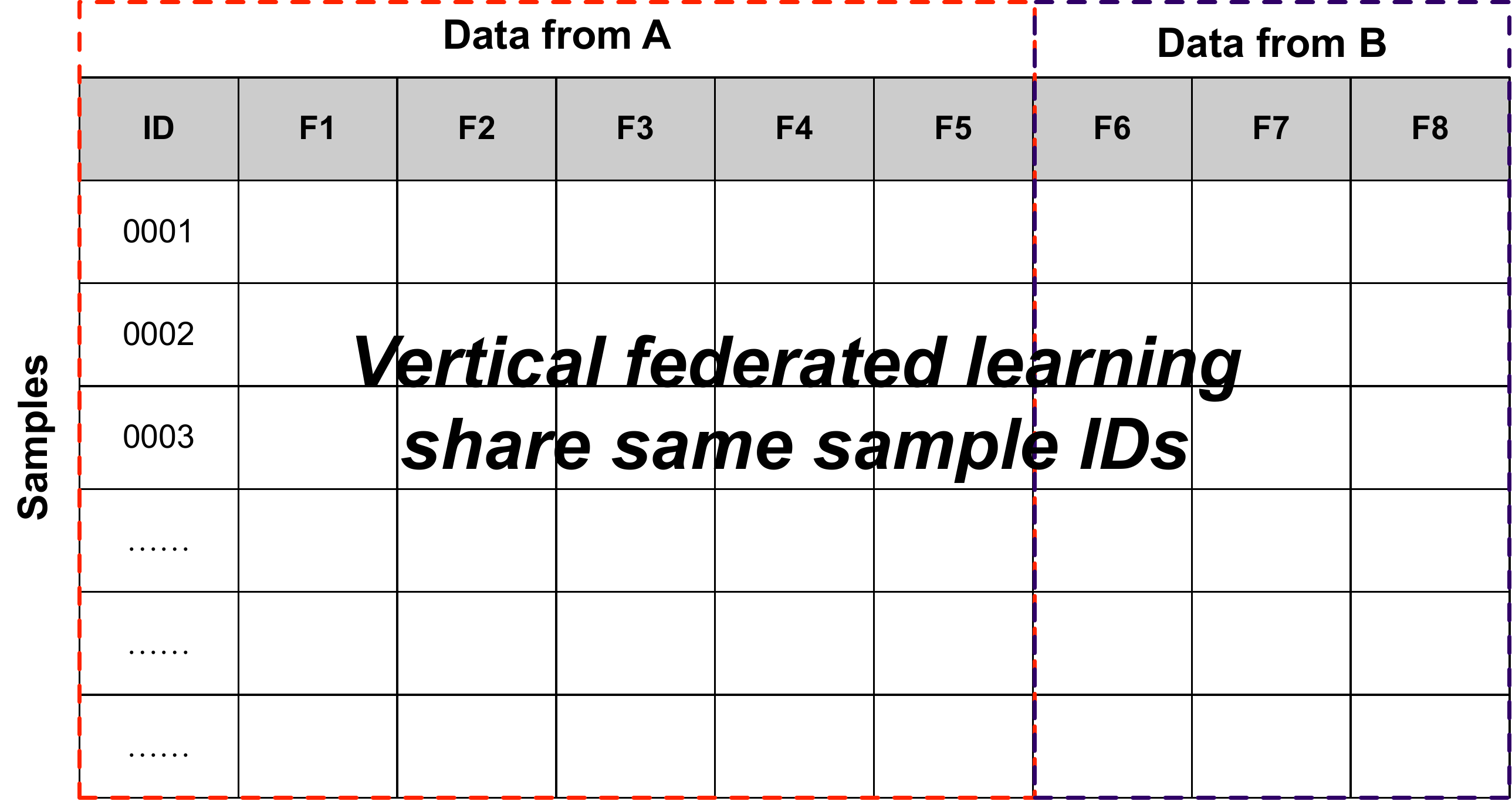}}
\caption{Types of Federated Learning: (a) horizontal federated learning and (b) vertical federated learning.}%(a) shows the data distribution of horizontal federated learning and (b) shows the distribution of vertical federated learning.}
\label{fig:federated_learning}
\end{figure*}

Based on the form of cooperation, FL can be further divided into two types: horizontal federated learning (HFL)~\cite{Bonawitz2019,bonawitzI2018ccs} and vertical federated learning (VFL)~\cite{yang2019acm,stephen2017corr}. HFL is designed for cooperation between participants who have the same feature space but different data samples. For example, the same type of companies in different cities (thus different clients/users). VFL is designed for participants who have the same sample IDs but different (even complementary) features. For example, two different types of companies in the same city (thus same clients/users), e.g., a bank and a commerce company.  Figure~\ref{fig:federated_learning} visualizes a two-participant example of their local data distributions in HFL and VFL. HFL and VFL work in different ways, follow different FL protocols, and adopt different privacy preservation techniques.

% \noindent \textbf{Vertical Federated Learning.} 
In this paper, we mainly focus on VFL since it has a broader scope of applications and is arguably more suitable for joint model training between large enterprises. For example, it is likely that a bank and an e-commerce company share a large amount of users (sample IDs) if they are in the same city. In this case, they can collaborate on the different user attributes they own using VFL to train a more powerful risk management model~\cite{yang2019acm}. VFL is more complex than HFL. Unlike in HFL, all participants share the same feature space where traditional learning algorithms still apply, in VFL, the participants own different features, thus require a new learning protocol for each machine learning algorithm.

A number of open-source frameworks have been developed by industry leaders to facilitate the application of FL in diverse and challenging real-world application scenarios. This include \texttt{PySyft}~\cite{pysyft} from OpenAI, \texttt{FATE}~\cite{fate} from WeBank, \texttt{TF Federated}~\cite{tffederated} from Google, \texttt{TF Encrypted}~\cite{tfencrypted} from Dropout and \texttt{CrypTen}~\cite{crypten} from Facebook. Particularly, \texttt{FATE}~\cite{fate}, \texttt{TF Encrypted} and \texttt{PySyft}~\cite{pysyft} are the three most popular real-world VFL frameworks. Meanwhile, for real-world applications, these frameworks often come with encryption-based collaboration protocols.  With the advancement of these open-source frameworks, FL has now become a mainstream choice for different data holders to collaboratively train higher performance models.  Such increasing popularity makes it more urgent to learn the privacy issue of FL protocols since even a negligible vulnerability could lead to data leakage of billions of samples.  In this work, we propose practical privacy attacks to examine the potential privacy risks of the privacy-preserving protocols implemented in those VFL frameworks. 

Many VFL protocols choose to sacrifice data security and user privacy for computing efficiency. We observe that \emph{such a sacrifice tends to make joint model training more vulnerable to the adversary, resulting in the leakage of private data.} Take one classic VFL protocol Privacy-preserving Logistic Regression~\cite{stephen2017corr,yang2019acm} as an example, it employs a third-party coordinator to coordinate and accelerate model training. The coordinator holds an RSA private key. It first sends the public key to the participants to encrypt their intermediate outputs, then combines the encrypted outputs uploaded by the participants, uses the private key to decrypt the combined outputs, and sends the decrypted result back to the participants. Undoubtedly, such a training procedure will leak too much private information to the coordinator. Even worse, in real-world scenarios, there is no guarantee that the coordinator will never deviate from the protocol specification or do not attempt to gather private information from the intermediate outputs.

Whilst several works have shown the possibility of inferring private data in privacy-preserving machine learning, they either focus on HFL~\cite{melis2019sp,nasr2019sp,hitaj2017ccs,wang2019infocom} or membership attacks \cite{nasr2019sp,pustozerova2020information} on adversarially secured models \cite{song2019spw}. 
Different to HFL, measuring privacy leakage in real-world VFL has two major challenges: 1) the intermediate outputs are encrypted; and 2) the participants' training data are heterogeneous. Encrypted intermediate outputs make it impossible for the participants to directly infer private data through gradient updates. As a consequence, existing learning-based attacks~\cite{zhu2019nips,melis2019sp,fredrikson2015ccs} become invalid in this setting. Moreover, since the participants’ training data are heterogeneous, none of them can infer private data though maliciously queries from the other.
%\texttt{TF Federated} provided by Google, \texttt{CrypTen} provided by Facebook, and \texttt{FATE} provided by WeBank, which have a difference in the encryption methods, protocols and even the architectures.

\noindent \textbf{Our work.} 
In this paper, we provide the first systematic study on the potential privacy risks in real-world VFL frameworks. Specially, our work seeks answers to two fundamental research questions: \emph{Is it possible to infer information about other participants' private data in real-word VFL protocols? If so, how much information can be inferred and how much computation it will cost?}

To answer the above questions, we study two widely used VFL algorithms: logistic regression~\cite{friedman2001elements} and XGBoost~\cite{chen2016kdd}. These two algorithms have been deployed in various applications by different organizations, including advertisement in e-commerce companies~\cite{zhou2017}, disease detection in hospitals~\cite{chen2017machine}, and fraud detection in financial companies~\cite{zhang2019distributed}.
We propose two simple but effective attacks: \emph{reverse multiplication attack} for logistic regression and \emph{reverse sum attack } for XGBoost, to investigate the possible training data leakage caused by numerical computations. In the reverse multiplication attack, the adversary attempts to steal the raw training data of a target participant by leveraging the intermediate multiplication results, when two participants are jointly training a logistic regression model. In the reserve sum attack, the adversary tries to probe the partial orders (key to tree-based learning) of the target participant's training data by inserting magic numbers in the exchanged data, when two participants are jointly training an XGBoost model. Though not completely equivalent to the raw data, the stolen partial orders can be further used to train an alternative model that is as effective as the one trained on the raw data.

We conduct the experiments in a real-world VFL framework \texttt{FATE}~\cite{fate}, and show that for both attacks, the adversary can infer various levels of the target participant's private data. 
% Specifically, to show the effectiveness of our reverse multiplication attack, we evaluate the privacy of participants' training data when they collaboratively train a logistic regression model on vertical partitions of the credit dataset. 
Specifically, the results of our \emph{reverse multiplication attack} on logistic regression VFL show that the learning protocol of logistic regression--which is claimed to be privacy-preserving to the training data--does not protect data privacy of the honest participant: via a collusion with the coordinator, the adversary can steal $100\%$ of the training data.  The results of the \emph{reverse sum attack } on XGBoost show that a malicious participant can perform accurate reverse sum attack against other participants. When collaboratively training a single boosting tree, the adversary can successfully steal the partial orders of samples with $100\%$ accuracy from the target participant.  Further,  the leaked information can be used to build an alternative classifier that is as accurate as the one trained on the original data, proving its potential commercial values.
% This shows that even well-designed learning protocols leak a significant amount of information about participants' training data, %and are 
% thus vulnerable to reverse multiplication attack.
These results indicate that, under reasonable threat models, even well-designed VFL protocols can leak a significant amount of information about the participants' private training data.

% Our experimental evaluation also shows that a malicious participant can perform accurate reverse sum attack against other participants. When collaboratively training a single boosting tree, the active participant can successfully steal partial orders of $\sim 3,200$ samples with $100\%$ accuracy from the target participant. We experimentally demonstrate that leaked information can be used to build such an alternative classifier that is as accurate as the one trained on the original data, showing the potential commercial values of the leaked leakage.

In summary, we make the following contributions:
\begin{itemize}
\item
 We conduct the first investigation on the potential privacy risks in vertical federated learning (VFL) via two proposed privacy attacks on real-world secure VFL protocols and frameworks,  including secure logistic regression and secure XGBoost.

\item 
We propose measures of data leakage caused by our two attacks, and show that an adversary (e.g.,  a semi-honest participant) with access to the learning process can infer different levels of the target participant's private data. We also conduct an in-depth analysis on the factors that influence data privacy.
\end{itemize}

This paper is organized as follows. In \S\ref{sec:preliminaries}, we introduce the formal definition of federated learning, the commonly used VFL protocols and the real-world frameworks for privacy preserving VFL. In \S\ref{sec:threat}, we formalize the privacy attacking problem and our threat model. \S\ref{sec:reverse_multiplication_attack} and \S\ref{sec:reverse_sum_attack} introduce the two proposed attacks, our implementations and the experimental results, separately. We discuss possible countermeasures in \S\ref{sec:countermeasure}, review related works in \S\ref{sec:related_work}, and finally conclude this paper in \S\ref{sec:conclusion}.

\section{Preliminaries}
\label{sec:preliminaries}

In this section, we introduce the mathematical notations, the security primitives used in VFL, two commonly used machine learning algorithms including logistic regression~\cite{friedman2001elements} and XGBoost~\cite{chen2016kdd}, and their privacy-preserving versions in VFL. We also briefly review several popular real-world VFL frameworks including \texttt{TF Federated},  \texttt{FATE}, \texttt{PySyft}, \texttt{TF Encrypted} and \texttt{CrypTen}.

\subsection{Notations} Here, we focus on the two-party VFL collaboration, the most extensively studied and practiced scenario in real-world VFL. 
Suppose there are two participants, $\mathcal{A}$ and $\mathcal{B}$, collaboratively training a distributed machine learning model following VFL. Let $\mathbf{X} \in \mathcal{R}^{n\times d}$ denote the \emph{complete dataset} containing $n$ samples, $\mathbf{Y} \in \mathcal{R}^{n\times1}$ denote the label set, and $\mathbf{I} \in \mathcal{R}^{n \times 1}$ denote the sample IDs. 
Participants $\mathcal{A}$ and $\mathcal{B}$ have the same sample IDs but own complimentary (non-overlapping) features: $\mathbf{X} = [\mathbf{X}_\mathcal{A} | \mathbf{X}_\mathcal{B}]$. The label set $\mathbf{Y}$ is held by $\mathcal{A}$. We also call $\mathcal{A}$ an \emph{activate participant} as it has the vital information of class labels and $\mathcal{B}$ a \emph{passive participant} as it only has the features. We further denote the $i$-th row of $\mathbf{X}$ as $\mathbf{x}_i$ (i.e., the $i$-th sample), and the part of $\mathbf{x}_i$ held by participant $\mathcal{A}$ as $\mathbf{x}_{i}^{\mathcal{A}}$ while the part of $\mathbf{x}_i$ held by $\mathcal{B}$ as $\mathbf{x}_i^{\mathcal{B}}$. That is, $\mathbf{x}_i = [\mathbf{x}_i^{\mathcal{A}} | \mathbf{x}_i^{\mathcal{B}}]$.

%\subsection{Federated Learning Protocols}

%\subsection{Federated Learning} 
%Federated learning aims to build machine learning models on datasets that are held by multiple participants without revealing anything but the model's output. Based on the characteristics of data distribution, the general federated learning can be categorized into \emph{horizontal federated learning}~\cite{Bonawitz2019,bonawitzI2018ccs} and \emph{vertical federated learning}~\cite{yang2019acm,stephen2017corr}. In horizontal federated learning, the distributed datasets share the same feature space but different samples. In contrast, in vertical federated learning, the distributed datasets share the same samples but differ in feature space. Figure~\ref{fig:federated_learning} visualizes the characteristics of data distribution in horizontal and vertical federated learning. Our work mainly evaluates the privacy leakage of learning a distributed machine learning model in vertical federated learning.
%
%\begin{figure*}[!th]
%\centering
%\subfloat[Horizontal Federated Learning]{\includegraphics[width=0.45\textwidth,height=0.2\textwidth]{fig/horizontal_federated_learning}} 
%\quad 
%\subfloat[Vertical Federated Learning]{\includegraphics[width=0.45\textwidth,height=0.2\textwidth]{fig/vertical_federated_learning}}
%\caption{Types of Federated Learning: (a) Horizontal federated learning and (b) Vertical federated learning.}%(a) shows the data distribution of horizontal federated learning and (b) shows the distribution of vertical federated learning.}
%\label{fig:federated_learning}
%\end{figure*}

\subsection{Security Primitives} 
% We review the primary privacy techniques adopted by vertical federated learning for providing privacy guarantees. 
VFL usually adopts additively homomorphic encryption like Paillier~\cite{paillier1999} to encrypt the intermediate outputs (e.g. gradients). Additively homomorphic encryption is a public key system that allows participants to encrypt their data with a known \emph{public key} and perform computation with the encrypted data by other participants with the same public key.  To extract the plaintext, the encrypted data needs to be sent to the private key holder for decryption. Let the encryption of a number $u$ be $[\![u]\!]$. For any plaintext $u$ and $v$, we have the following additive operation,
\begin{equation}
[\![u+v]\!] = [\![u]\!] + [\![v]\!].
\end{equation}
We can also multiply a ciphertext with a plaintext as follows,
\begin{equation}
[\![v\cdot u]\!] =  v\cdot [\![u]\!],
\end{equation}
where $v$ is not encrypted.
These operations can be extended to work with vectors and matrices component-wise. For example, the inner product between a plain vector $\mathbf{v}$ and a cipher vector $[\![\mathbf{u}]\!]$: $\mathbf{v}^\text{T}\cdot [\![\mathbf{u}]\!] = [\![\mathbf{v}^\text{T} \cdot \mathbf{u}]\!]$. 

\subsection{VFL Algorithms and Protocols}
\label{sec:learning_protocol}

Logistic regression~\cite{friedman2001elements} and XGBoost~\cite{chen2016kdd} are arguably the two most widely used machine learning algorithms. They constantly appear in the winning solutions of Kaggle\footnote{\url{https://www.kaggle.com/competitions}} competitions~\cite{xgboostwinner}. 
Due to their simplicity and efficiency, these two algorithms have also become the most classic VFL algorithms in many VFL frameworks.
% These two algorithms are also the most classic frequently used by giant organizations for collaboratively model training using FL.
% For example, the financial institution and  the e-commerce company use linear regression to collaboratively train a scorecard model on their own private training data \cite{xxxx}.
Here, we introduce the main ideas of linear regression and XGBoost, and their privacy-preserving versions in VFL.

\subsubsection{Logistic Regression}
Given $n$ training samples $\{\mathbf{x}_i\}^{i=1\cdots n}$ with $d$ features ($\mathbf{x}_i \in \mathbb{R}^d$), and $n$ corresponding output labels $\{y_i\}^{i=1\cdots n}$, linear regression is to learn a mapping function such that $f(\mathbf{x}_i) = \mathbf{y}_i, \forall \mathbf{x}_i$.  The function $f$ is assumed to be linear and can be written as the inner product between $\mathbf{x}_i$ and a $d$-dimensional coefficient vector $\boldsymbol{\theta}$:
$
f(\mathbf{x}_i) = \sum_{j=1}^{d} \boldsymbol{\theta}^j\mathbf{x}_{i}^{j}  = \boldsymbol{\theta}\mathbf{x}_i.
$
  The coefficient $\boldsymbol{\theta}$ can be learned by minimizing the empirical error on all training samples defined by the following loss function,
\begin{equation}
\label{eq:lr}
 L(\boldsymbol{\theta}) = \frac{1}{n} \sum_{i=1}^{n} \frac{1}{2} {(\boldsymbol{\theta}{\mathbf{x}_i} - \mathbf{y}_i)}^2.
\end{equation}
In traditional centralized machine learning, the above minimization problem can be solved efficiently by the Stochastic Gradient Descent (SGD)~\cite{sgd} optimizer via iterative minibatch training. 
% The above objective function is convex so SGD can converge to the global optimal.
% is an effective approximation algorithm for approaching a local minimum of a function, step by step.

\noindent \textbf{Loss decomposition.} Different from centralized learning,  each $\mathbf{x}_i$ in VFL is partitioned into $[\mathbf{x}_i^\mathcal{A}|\mathbf{x}_i^\mathcal{B}]$. Accordingly, the coefficient $\boldsymbol{\theta}$ will be partitioned into: $[\boldsymbol{\theta}^\mathcal{A} | \boldsymbol{\theta}^\mathcal{B}]$.  The gradient of the loss function of Eq.~(\ref{eq:lr}) can then be decomposed as, 
\begin{align}
\Delta L(\boldsymbol{\theta}) &\approx \frac{1}{n} \sum_{i=1}^{n} (\frac{1}{4}\boldsymbol{\theta}^\mathcal{A} \mathbf{x}_i^\mathcal{A}+ \frac{1}{4} \boldsymbol{\theta}^\mathcal{B} \mathbf{x}_i^\mathcal{B} - \frac{1}{2}\mathbf{y}_i) \cdot [\mathbf{x}_i^{\mathcal{A}}|\mathbf{x}_i^{\mathcal{B}}] \\
 & = \frac{1}{n} (\frac{1}{4}\boldsymbol{\theta}^{\mathcal{A}}\mathbf{X}^{\mathcal{A}} + \frac{1}{4}\boldsymbol{\theta}^{\mathcal{B}}\mathbf{X}^{\mathcal{B}} - \frac{1}{2}\mathbf{Y}) \cdot [\mathbf{X}^\mathcal{A}|\mathbf{X}^{\mathcal{B}}]
 \label{eq:vertical_loss}
\end{align}
Participants $\mathcal{A}$ and $\mathcal{B}$ first compute their corresponding terms in the above decomposition on their private training data, then exchange their outputs with or without a third-party coordinator to form the complete output.
In \cite{stephen2017corr, yang2019acm}, a third-party coordinator is involved to accelerate the output exchange and the training process. Next, we will show the detailed steps of computing the above terms in VFL.

\begin{table}[!th]
\centering
\caption{Steps of two-party logistic regression in VFL.}
\resizebox{.45\textwidth}{!}{
\begin{tabular}{p{0.01cm} |p{2.1cm} |p{2,5cm} |p{3.5cm} } 
\hline  & \textbf{Participant} $\mathcal{A}$ & \textbf{Participant} $\mathcal{B}$ & \textbf{Coordinator} $\mathcal{C}$ \\ 
\hline 
\hline 
1 & init $\boldsymbol{\theta}^\mathcal{A};$ & init $\boldsymbol{\theta}^\mathcal{B}$; & send the public key to $\mathcal{A}$ and $\mathcal{B}$;\\
%2 & select the next batch indices $S$;  & &\\
2 & compute $[\![\mathbf{u}]\!] = [\![\frac{1}{4}\boldsymbol{\theta}^\mathcal{A}\mathbf{X}_{S}^\mathcal{A}-\frac{1}{2}\mathbf{Y}_{S}]\!]$ and send to $\mathcal{B}$; 
& a) compute $[\![\mathbf{v}]\!] =  [\![\frac{1}{4} {\boldsymbol{\theta}}^{\mathcal{B}}\mathbf{X}_{S}^{\mathcal{B}}]\!] + [\![\mathbf{u}]\!] $ and send  to  $\mathcal{A}$; \newline
   b) compute $[\![\mathbf{v}]\!]  \mathbf{X}_S^{\mathcal{B}}$ and sent to  $\mathcal{C}$; \\
3  & compute $[\![\mathbf{v}]\!]  \mathbf{X}_S^{\mathcal{A}}$ and sent to  $\mathcal{C}$; &&\\
4 & & & decrypt  $[\![\mathbf{v}]\!]  \mathbf{X}_{S}^{\mathcal{A}}$ and $[\![\mathbf{v}]\!]  \mathbf{X}_S^{\mathcal{B}}$ to get the gradient $g^{\mathcal{A}}, g^{\mathcal{B}}$, and send them to $\mathcal{A}$ and $\mathcal{B}$.\\
\hline
\end{tabular}}
\label{table:logistic_regression}
\end{table}

%In Table~\ref{table:logistic_regression},  $S$ denotes the  indices of a mini-batch examples. As shown in Table~\ref{table:logistic_regression}, participants A and B first initialize the distributed coefficient $\boldsymbol{\theta}^\text{A}$ and $\boldsymbol{\theta}^\text{B}$, and the coordinator C send the public key to A and B. Then, A and B collaboratively compute the distributed loss function of Eq.~(\ref{eq:vertical_loss}) on a mini-batch $S$. Finally, the third-party coordinator C forms the final gradient calculated  on $S$,  and sends them back to A and B, respectively.

\noindent \textbf{Secure logistic regression.} To compute Eq.~(\ref{eq:vertical_loss}) without exposing the private data,  most secure VFL protocols adopt additively homomorphic encryption like Paillier to encrypt the intermediate outputs and operate on the ciphertext~\cite{mohassel2017sp,yang2019acm,stephen2017corr}. 
The encrypted distributed gradient of Eq.~(\ref{eq:vertical_loss}) is 
\begin{equation}
[\![\Delta L(\boldsymbol{\theta}) ]\!] \approx 
 \frac{1}{n} ([\![\frac{1}{4}\boldsymbol{\theta}^{\mathcal{A}}\mathbf{X}^{\mathcal{A}} ]\!] + [\![\frac{1}{4}\boldsymbol{\theta}^{\mathcal{B}}\mathbf{X}^{\mathcal{B}}]\!] - [\![\frac{1}{2}\mathbf{Y}]\!]) \cdot [\mathbf{X}^\mathcal{A}|\mathbf{X}^{\mathcal{B}}]
\label{eq:encrypted_loss}
\end{equation} 
In VFL, we also use the SGD optimizer to update the model parameters. Table~\ref{table:logistic_regression} lists the detailed training steps of logistic regression in secure VFL. 

The secure VFL protocol works as follows.
First, participants $\mathcal{A}$ and $\mathcal{B}$ initialize $\boldsymbol{\theta}^{\mathcal{A}}$ and $\boldsymbol{\theta}^{\mathcal{B}}$ as vectors of random values or all $0$s. Coordinator $\mathcal{C}$ creates an encryption key pair and sends the public key to $\mathcal{A}$ and $\mathcal{B}$. Second, in each training iteration (communication round), participant $\mathcal{A}$ randomly selects a mini-batch of indices $S$ and updates the distributed coefficient by averaging partial derivatives of all samples to the current coefficient. Steps 2--4 in Table~\ref{table:logistic_regression} describe how $\mathcal{A}$ and $\mathcal{B}$ collaboratively compute the gradient of the encrypted distributed loss function and update the distributed coefficient. I.e., using an additively homomorphic encryption scheme, participants $\mathcal{A}$ and $\mathcal{B}$ compute on the encrypted intermediate outputs and generate an encrypted derivative which, when decrypted, matches with the derivative as if it had been calculated on the plaintext. 
% We refer readers to \cite{mohassel2017sp,yang2019acm,stephen2017corr} for more details about secure logistic regression.

% For the update, using an additively homomorphic encryption scheme, participants A and B can compute on the encrypted intermediate outputs, generating an encrypted derivative which, when decrypted, matches to the derivative as if it had been calculated on the plaintext. 
%Specifically,   A  computes  $\frac{1}{4}[[\mathbf{x}_i^\text{A} \cdot \mathbf{w}^\text{A}]] - [[\frac{1}{2}y_i ]]$ and B computes $\frac{1}{4} [[\mathbf{x}_i^\text{B}\cdot \mathbf{w}^\text{B}]]$, and then they  communicate to get the ciphertext $\frac{1}{4}[[\mathbf{x}_i^\text{A} \cdot \mathbf{w}^\text{A}]] + \frac{1}{4} [[\mathbf{x}_i^\text{B}\cdot \mathbf{w}^\text{B}]] -  [[\frac{1}{2}y_i ]]$ which is the left term of Equation (\ref{eq:encrypted_loss}). In the end, A computes $(\frac{1}{4}[[\mathbf{x}_i^\text{A} \cdot \mathbf{w}^\text{A}]] + \frac{1}{4} [[\mathbf{x}_i^\text{B}\cdot \mathbf{w}^\text{B}]] -  [[\frac{1}{2}y_i ]])\cdot \mathbf{x}_i^\text{A}$,  B computes $(\frac{1}{4}[[\mathbf{x}_i^\text{A} \cdot \mathbf{w}^\text{A}]] + \frac{1}{4} [[\mathbf{x}_i^\text{B}\cdot \mathbf{w}^\text{B}]] -  [[\frac{1}{2}y_i ]])\cdot \mathbf{x}_i^\text{B}$, and then they communicate again to obtain  the whole ciphertext. Refer \cite{mohassel2017sp,yang2019acm,stephen2017corr} for more details about secure logistic regression.

\subsubsection{XGBoost} Given $n$ training samples $\{\mathbf{x}_i\}^{i=1\cdots n}$ with $d$ features ($\mathbf{x}_i \in \mathbb{R}^d$), and $n$ corresponding output labels $\{y_i\}^{i=1\cdots n}$, XGBoost learns a function consisting of $k$ regression trees such that  $f(\mathbf{x}_i) = \sum_{t=1}^{k} f_t(\mathbf{x}_i), \forall \mathbf{x}_i$~\cite{chen2016kdd}. To learn such $k$ regression trees, XGBoost adds a tree $f_t$ at the $t$-th iteration to minimize the following loss,
\begin{equation}
L^{(t)} \approx \sum_{i=1}^{n} [l(y_i, \hat{y}^{(t-1)}) + g_i f_t(\mathbf{x}_i) + \frac{1}{2}h_i{f}_t^2(\mathbf{x}_i)] + \Omega(f_t),
\end{equation}
where $g_i$ and $h_i$ are the first and second order gradient statistics of the loss function calculated by $f_{t-1}$, and $\Omega$ is a regularization term reducing the complexity of the model.

To construct the $t$-th regression tree,  XGBoost starts from a leaf node and iteratively adds optimal branches to the tree until the terminating condition is reached. The loss reduction after the split is,
\begin{equation}
\label{eq:reduction_loss}
L_{split} = \frac{1}{2}[\frac{{(\sum_{i\in I_{L}}g_i)}^2}{\sum_{i\in I_L}h_i + \lambda} + \frac{{(\sum_{i\in I_{R}}g_i)}^2}{\sum_{i\in I_R}h_i + \lambda} -\frac{{(\sum_{i\in I}g_i)}^2}{\sum_{i\in I}h_i + \lambda} ] - \gamma,
\end{equation}
where $I_L$ and $I_R$ are the instance space of the left and right nodes after the split, respectively.  In practice, XGBoost chooses an optimal split that can maximize the above reduction loss. That is, in each iteration, XGBoost enumerates all the possible split features and values, then finds the one that maximizes the reduction loss. 
% We refer readers to~\cite{chen2016kdd} for more details about XGBoost.

\noindent \textbf{Secure XGBoost.}  Secure XGBoost~\cite{secureboost}, also known as SecureBoost, is the privacy-preserving version of XGBoost for secure VFL. SecureBoost allows the learning process of $k$ regression trees to be jointly conducted between two participants with shared sample IDs but different features. To learn a privacy-preserving regression tree, SecureBoost starts from a single leaf node and coordinates two computing participants to find the optimal split.

Different from secure logistic regression, SecureBoost does not rely on a third-party coordinator to coordinate the learning process.  In SecureBoost, the participant who holds the label information serves as a coordinator. We also call the participant who holds the labels as the activate participant while the other as the passive participant.

In SecureBoost, the activate and passive participants collaboratively calculate the reduction loss of Eq.~(\ref{eq:reduction_loss}) and search for the best split. In each iteration, it operates as follows. \textbf{First}, the activate participant calculates the first and second order gradients for all training samples, then encrypts and sends the encrypted gradients to the passive participant. \textbf{Second}, the passive participant receives the encrypted gradients and calculates the reduction loss for all possible split features and values. For efficiency and privacy, the passive participant maps the features into data bins and aggregates the encrypted gradient statistics based on the bins, rather than directly working on $[\![g_i]\!]$ and $[\![h_i]\!]$. \textbf{Third}, the activate participant receives and decrypts the encrypted reduction loss,  and then finds an optimal split dimension and feature having the maximum reduction. 
% Refer to~\cite{secureboost} for more details about SecureBoost.

\subsection{Summary of Real-world VFL Frameworks}
\begin{table}[!th]
\centering
\caption{Popular real-world secure FL frameworks. HE: homomorphic encryption~\cite{homomorphicencryption}; SS: Secret Sharing~\cite{secretsharing}.}
\resizebox{.45\textwidth}{!}{
\begin{tabular}{c|c|c|c|c} 
\hline  \textbf{Framework} &  \textbf{Developer} & \textbf{Vertical} & \textbf{Horizontal } & \textbf{Encryption} \\
\hline \hline
 \texttt{FATE} & WeBank& \cmark  & \cmark  & HE \\ 
\texttt{PySyft} & OpenAI &  \cmark  & \cmark & HE, SS  \\ 
\texttt{TF Federated} & Google & \xmark & \cmark  & SS  \\ 
\texttt{TF Encrypted} & Dropout&  \cmark  &  \cmark &HE, SS   \\ 
\texttt{CrypTen} & Facebook& \cmark & \cmark  & HE, SS   \\ 
\hline
\end{tabular}}
\label{table:library}
\end{table}

Table~\ref{table:library} summarizes the functionalities and basic encryption techniques supported by several popular real-world FL frameworks, including \texttt{PySyft}~\cite{pysyft}, \texttt{FATE}~\cite{fate}, \texttt{TF Federated}~\cite{tffederated}, \texttt{TF Encrypted}~\cite{tfencrypted} and \texttt{CrypTen}~\cite{crypten}. All of these frameworks except \texttt{TF Federated} support VFL protocols. For instance, \texttt{FATE} provides VFL protocols of logistic regression and tree-based learning; \texttt{PySyft} provides basic modules (e.g., communication and encrypted computation modules) for the fast implementation of VFL protocols.

% Of the five evaluated frameworks, \texttt{FATE}, \texttt{PySyft}, \texttt{TF Encrypted} and \texttt{CrypTen} support participants to train a joint model when their training data is vertically aligned or horizontally aligned. The remaining one \texttt{TF Federated} only supports to training a model when the data are horizontally aligned. For the encryption techniques, these five frameworks mainly utilize two mainstream methods, homomorphic encryption~\cite{homomorphicencryption} and secret sharing~\cite{secretsharing}.

Next, we will introduce our proposed attacks for secure logistic regression and SecureBoost separately, and evaluate the attacks within the \texttt{FATE} and \texttt{PySyft} frameworks. Note that our attacks are designed to attack the fundamental VFL protocols rather than one particular VFL framework.

\section{Threat Model}
\label{sec:threat}
%Although secure learning protocols and open-source frameworks provide the possibility for different organizations to collaboratively train a machine learning model, we find that those protocols' over pursuit for learning efficiency, however, lead to the privacy risk of participants' training data. By constructing two novel attacks, \emph{reverse multiplication attack} and \emph{reverse sum attack}, we show there indeed exists vulnerabilities against vertical federated learning, and an adversary can steal the plaintext of participants' private data. In this section, we present the threat model and privacy risks faced by vertical federated learning.

%Figure~\ref{fig:federated_learning_pipeline} shows a typical federated learning scenario.
We consider the two-party VFL and a semi-honest adversary who merely cooperates for gathering information about the other participant's private data out of the learning process, but does not deviate from the protocol specification. 
In other words, the adversary attempts to steal the other participant's data without sabotaging the learning protocol.
%In the following, we describe the threat model in vertical federated learning.
We investigate whether this is achievable even when the VFL protocol is secured using additively homomorphic encryption.
 
\begin{figure}[!ht]
\centering
\includegraphics[width=0.48\textwidth]{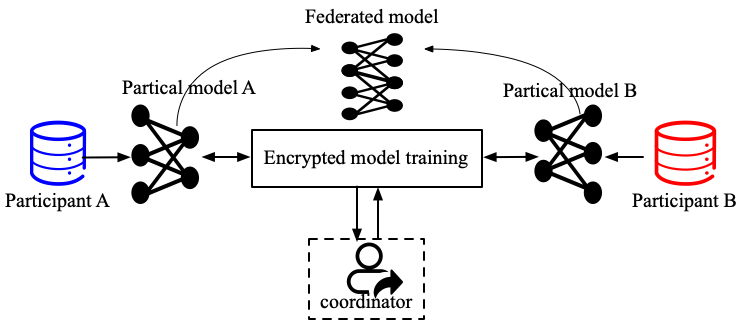}
\caption{A typical two-party VFL scenario. The complete dataset  $\mathbf{X}$  is vertically split into $\mathbf{X}_\mathcal{A}$  and $\mathbf{X}_\mathcal{B}$ and held by participants  $\mathcal{A}$ and $\mathcal{B}$ respectively. Participants $\mathcal{A}$ and $\mathcal{B}$ maintain partial models locally and exchange the encrypted partial gradient and other essential information to jointly learn a global model on the complete dataset $\mathbf{X}$.}
\label{fig:federated_learning_pipeline}
\end{figure}

%\subsection{Threat Model}
%\label{sec:threat_model}
Figure~\ref{fig:federated_learning_pipeline} illustrates a typical two-party VFL scenario where the two participants ($\mathcal{A}$ and $\mathcal{B}$) keep their private data locally and jointly train a global model by sharing only the encrypted partial gradient.  In secure logistics regression, $\mathcal{A}$ and $\mathcal{B}$ require a third-party coordinator to handle the encrypted partial gradients and update the global model. In SecureBoost, the two participants independently learn a joint model without any third-party.  
% We assume that the adversarial participant is semi-honest that does not deviate from the protocol specification, but try to gather as much private information out of the learning protocol as possible. More details about the  assumption are listed below.

%We investigate \textbf{\textit{if an adversarial participant can steal the privacy about other participant's training data, even when the learning protocol is privacy-preserving.}}  As shown in Figure~\ref{fig:federated_learning_pipeline}, participants $\mathcal{A}$ and $\mathcal{B}$ keep their private data locally, and according to a particular secure learning protocol, they jointly train a model without revealing anything but the model's output. In some learning protocols like logistics regression, $\mathcal{A}$ and $\mathcal{B}$ require a third-party coordinator to handle the encrypted partial gradients and update the global model. In contrast, some other protocols like SecureBoost, the two participants, can independently learn a joint model without any third-parties. In both types of learning protocols, we assume that the adversarial participant is semi-honest that does not deviate from the protocol specification, but try to gather as much private information out of the learning protocol as possible. 

%{\color{red}TODO: When collaboratively training a joint model, the adversarial participant stealthily.}
\vspace{0.05in}
\noindent \textbf{\textbf{Adversary.}} We assume that one of the two participants is the adversary who can send or receive the encrypted information (e.g., gradient and the multiplication result between gradient and data) of the other participant's training data. 
%In general, the adversary maliciously controls the local training process. 
% This curious but semi-honest adversary merely operates to gather as much private information out of the learning protocol as possible. 
 In our first attack the \emph{reverse multiplication attack} (refer \S\ref{sec:reverse_multiplication_attack} for more details) proposed for secure logistic regression, the adversary may also corrupt the third-party coordinator to gather more private information out of the protocol.
 In our second attack the \emph{reverse sum attack} (refer \S\ref{sec:reverse_sum_attack} for more details) proposed for SecureBoost, the adversary may construct specific strings and insert them into the least significant bits of gradients while keeping the learning protocol intact.

\vspace{0.05in}
\noindent \textbf{Adversary's objectives.} The adversary's main objective is to infer as much the other participant's private data as possible. In our reverse multiplication attack, the adversary aims to infer the target participant's raw training data. In our reverse sum attack, the adversary attempts to infer the partial orders of the target participant's training samples.

\vspace{0.05in}
\noindent \textbf{Assumption about the third-party coordinator.} In our reverse multiplication attack, we assume the adversary could corrupt the third-party coordinator. This assumption is reasonable since finding a trusted third-party coordinator is extremely difficult. In practice,  for convenience, the third-party coordinator is usually deployed and maintained by one of the participant. For example, a financial institution will probably serve as the coordinator in its collaboration with an e-commerce company. On the other hand, the third-party coordinator also has strong monetary incentives to collude with the adversary, considering the high value of the gathered heterogeneous data.
% when a financial institution collabarete with an e-commerce company for jointly training a model. This financial institution may probably deploy the third-party coordinator locally and jointly train the model with the e-commerce company.  
% In this case, it is easy for the adversarial participant to compromise the coordinator. 
Even if the coordinator is protected by TEEs (Trusted Execution Environments)~\cite{tee} like SGX~\cite{sgx}, it still could be comprised by various state-of-the-art attacks, e.g., SGX has been proven to be vulnerable to side-channel attacks~\cite{wang2017ccs}.
% {\color{red}{Even worse, the third-party coordinator is deployed Ant Group}}

%Like the controlled participant, we assume that the semi-honest adversary does not destroy the third-party's protocol specification. This semi-honest adversary only exploits the limited statists that the corrupted third-party received or other utilities held by this third-party. Usually, in vertical federated learning, the third-party coordinator concatenates the encrypted intermediate outputs received from the two participants and form the final gradient updates. Thus, the adversary could use such gradient updates to infer the target participant's training data. In $\S$\ref{sec:reverse_multiplication_attack}, we will illustrate how the adversary reverse-engineers the multiplication terms of the training data leveraging the third-party coordinator.

\section{Reverse Multiplication Attack}
\label{sec:reverse_multiplication_attack}
This reverse multiplication attack is designed against secure logistic regression based VFL \cite{mohassel2017sp,yang2019acm,stephen2017corr}. 
In this attack, the adversary's goal is to reverse-engineer each multiplication term of the matrix product (see Eq. \eqref{eq:encrypted_loss}), so as to infer the target participant's raw training data. Note matrix multiplication is one fundamental operation in many FL protocols. 
% Thus, we focus on logistic regression in a vertical federated setting and propose a simple yet effective reverse multiplication attack against it for inferring the target participant's raw training data.
Figure~\ref{fig:reverse_multiplication_attack_pipeline} illustrates the attack pipeline. We implement the reverse multiplication attack on one popular VFL framework, \texttt{FATE}, and empirically measure the data (privacy) leakage raised by matrix multiplication through performing the reverse multiplication attack.

\subsection{Attack Pipeline}
\label{sec:collusion_attack}

\begin{figure}[!ht]
\centering
\includegraphics[width=0.48\textwidth]{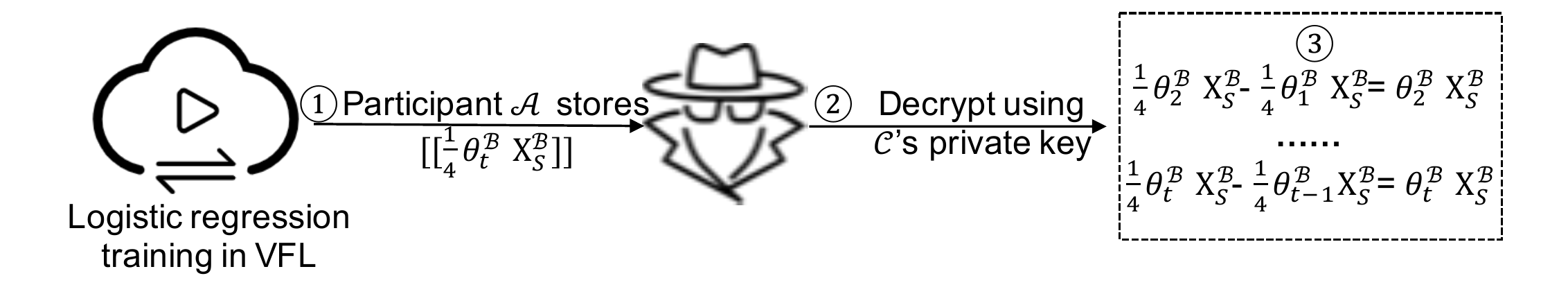}
\caption{Pipeline of the reverse multiplication attack.}
\label{fig:reverse_multiplication_attack_pipeline}
\end{figure}

% As shown in Table~\ref{table:logistic_regression}, logistic regression in vertical federated learning involves both computing participants and a third-party coordinator. 
% In our work, we focus on the simple scenario where only two data holders, $\mathcal{A}$ and $\mathcal{B}$,  participant in the model training process. 

Logistics regression relies on a series of multiplications between coefficients and participants' private training data to deduce the gradient, as decomposed in Eq.~(\ref{eq:encrypted_loss}). Participants $\mathcal{A}$ and $\mathcal{B}$ are required to exchange the intermediate multiplication results for calculating the gradient. Since the multiplication result is encrypted by the public key and contains no information about the training data, the adversarial needs to collude with the private key owner (the coordinator) to steal raw value of the encrypted multiplication result. Once the raw multiplication result is obtained, we can leverage linear algebra and scientific computation to reverse-engineer each multiplication terms (i.e., raw training data) from the multiplication result. 

Suppose participate $\mathcal{A}$ is the adversary and coordinator $\mathcal{C}$ is corrupted.  Participate $\mathcal{A}$ attacks $\mathcal{B}$ using the following information: 1) the intermediate output $[\![\frac{1}{4} {\boldsymbol{\theta}}_{t}^{\mathcal{B}}\mathbf{X}_{S}^{\mathcal{B}}]\!]$ received from $\mathcal{B}$; and 2) the partial gradient $g_{t}^{\mathcal{B}}$ that $\mathcal{C}$ sends back to $\mathcal{B}$. The intermediate output is the linear product of $\mathcal{B}$'s partial gradient and features. Taking $\mathcal{B}$'s features as unknown parameters, participant $\mathcal{A}$ can mount the reverse multiplication attack by solving the following linear functions,
\begin{equation} \label{eq1}
\begin{split}
 \frac{1}{4} {\boldsymbol{\theta}}_{1}^{\mathcal{B}}\mathbf{X}_{S}^{\mathcal{B}} - \frac{1}{4} {\boldsymbol{\theta}}_{0}^{\mathcal{B}}\mathbf{X}_{S}^{\mathcal{B}} &= g_{1}^{\mathcal{B}}\mathbf{X}_{S}^{\mathcal{B}} \\
 \frac{1}{4} {\boldsymbol{\theta}}_{2}^{\mathcal{B}}\mathbf{X}_{S}^{\mathcal{B}} - \frac{1}{4} {\boldsymbol{\theta}}_{1}^{\mathcal{B}}\mathbf{X}_{S}^{\mathcal{B}} &= g_{2}^{\mathcal{B}}\mathbf{X}_{S}^{\mathcal{B}} \\
 &\cdots \\
  \frac{1}{4} {\boldsymbol{\theta}}_{t}^{\mathcal{B}}\mathbf{X}_{S}^{\mathcal{B}} - \frac{1}{4} {\boldsymbol{\theta}}_{t-1}^{\mathcal{B}}\mathbf{X}_{S}^{\mathcal{B}} &= g_{t}^{\mathcal{B}}\mathbf{X}_{S}^{\mathcal{B}} \\
   &\cdots 
\end{split}
\end{equation}

Figure \ref{fig:reverse_multiplication_attack_pipeline}  illustrates the detailed steps of the reverse multiplication attack. The three attack steps are as follows.
% We consider that the data samples held by $\mathcal{A}$ and $\mathcal{B}$ are already aligned.

\begin{itemize}
%\item \textbf{Step 1}: 
%The adversary chooses $m$ object examples, and tries to infer the features held by B.   Let $\mathbf{I}=\{\mathbf{i}_j | j=1, 2, \cdots m\}$ denote the indices of these examples. 

\item \textbf{Step 1.}  The adversary stealthily stores the intermediate encrypted multiplication result computed on each mini-batch. Specifically, for each mini-batch $S$, the adversary stores ($\romannumeral1$) the multiplication result between $\mathcal{B}$'s partial data and coefficient, $[\![\frac{1}{4} {\boldsymbol{\theta}}_{t}^{\mathcal{B}}\mathbf{X}_{S}^{\mathcal{B}}]\!]$, and ($\romannumeral2$) the coefficient update of gradient, $g_{t}^{\mathcal{B}}$. 
Note that the encrypted product is shared by $\mathcal{B}$ for jointly calculating the encrypted gradient on the mini-batch.

%\item \textbf{Step 2}: The adversary manipulates \texttt{A} to generate a mini-batch $\mathbf{s}$ containing the selected  indices $\mathbf{I}$.
\item \textbf{Step 2.} The adversary decrypts $[\![\frac{1}{4} {\boldsymbol{\theta}}_{t}^{\mathcal{B}}\mathbf{X}_{S}^{\mathcal{B}}]\!]$ calculated on each mini-batch. Note that $\mathcal{A}$ and $\mathcal{B}$ use the same public key for encryption, and $\mathcal{C}$ uses the private key for decryption.  Hence, the adversary can use $\mathcal{C}$'s private key to decrypt the ciphertext.

\item \textbf{Step 3.} The adversary reverses the multiplication terms to obtain $\mathcal{B}$'s training data.  For a specific mini-batch $\mathbf{X}_{S}^{\mathcal{B}}$, the adversary has the following equation on two successive training epochs: $\frac{1}{4} {\boldsymbol{\theta}}_{t}^{\mathcal{B}}\mathbf{X}_{S}^{\mathcal{B}} - \frac{1}{4} {\boldsymbol{\theta}}_{t-1}^{\mathcal{B}}\mathbf{X}_{S}^{\mathcal{B}} =g_{t}^{\text{B}}\mathbf{X}_{S}^{\mathcal{B}}$, where only  $\mathbf{X}_{S}^{\mathcal{B}}$ is the unknown parameter. \emph{The adversary can compute the unknown} $\mathbf{X}_{S}^\mathcal{B}$ \emph{if it gets sufficient equations.} It requires at least $|\mathcal{B}|$ equations to solve the unknown parameter, where $|\mathcal{B}|$ denotes the number of features in $\mathcal{B}$'s private data. 

\end{itemize}

The above steps describe how the adversary utilizes the intermediate output  $[\![\frac{1}{4} {\boldsymbol{\theta}}_{t}^{\mathcal{B}}\mathbf{X}_{S}^{\mathcal{B}}]\!]$, and partial gradient $g^\mathcal{B}$ to steal $\mathcal{B}$'s private training data.  In some VFL frameworks, the coordinator directly updates the coefficient and sends it back to the participants. In this case, the adversary can solve the unknown parameter $\mathbf{X}_{S}^\mathcal{B}$ simply using  several multiplication results $\boldsymbol{\theta}_{t}^{\mathcal{B}} \mathbf{X}_{S}^\mathcal{B}$ and the partial gradients $\boldsymbol{\theta}_{t}^{\mathcal{B}}$.

The success of reverse multiplication attack depends on the rank\footnote{The rank of a matrix is defined as the maximum number of linearly independent column vectors or row vectors in the matrix.} of the coefficient matrices of the equations. The adversary can steal the full private data if the coefficient matrix is fully ranked, otherwise, the rank can be used to quantitatively measure how much data will be leaked. In the experiments, we first explore the influence of training parameters on the rank of the coefficient matrix, then apply our reverse multiplication to attack secure logistic regression. 
%  implemented using four real-world PPML frameworks, including \texttt{FATE}, \texttt{PySyft} and \texttt{TF Encrypted}, and \texttt{CrypTen}.

\subsection{Experimental Setup}
\label{sec:experiment_multiplication}

\noindent\textbf{Datasets.} 
We use three classic classification datasets:  Credit~\cite{ucicredit},  Breast~\cite{ucibreast}, and Vehicle~\cite{ucivehicle} to estimate the performance of the reverse multiplication attack.
\begin{itemize}
\item Credit  contains the payment records of $30,000$ customers from a bank, where $~5,000$ customers are malicious while the rest of the customers are all benign. Each record has a total of $23$ integer or real-valued features. The Credit dataset is a popular benchmark dataset used to evaluate binary classification tasks~\cite{ye2009eswa}.
\item  Breast is also a binary classification dataset. It contains $699$ medical diagnosis records of breast cancer, of which $357$ are benign while the rest $212$ are malignant. Each record has $30$ features extracted from a digitized image of a fine needle aspirate of a breast mass.
\item Vehicle is a multi-classification dataset extracted from example silhouettes of objects. 
 It contains $946$ samples belonging to four vehicle categories. Each sample has a set of $18$ features extracted from the silhouette of the original object.

\end{itemize}

\noindent\textbf{Setup for secure logistic regression.}
We implement the secure logistic regression protocol in the most popular real-world secure and private VFL framework, \texttt{FATE}. % and \texttt{PySyft}. 
% We then test our reverse multiplication attack against these two implementations.
For training, we consider two participants $\mathcal{A}$ and $\mathcal{B}$ along with a third-party coordinator $\mathcal{C}$. Detailed implementations are as follows.

\texttt{FATE} provides many default implementations of machine learning algorithms, including logistic regression and tree-based algorithms. In our experiment, we use \texttt{FATE}'s default implementation of privacy-preserving logistic regression~\cite{yang2019acm}. This implementation employs a third-party arbiter to coordinate the process of gradient computation and send back the decrypted gradient to the participants. For data privacy protection, this implementation uses an RSA private key to encrypt the intermediate output which can only be decrypted by the public key owner (i.e., the arbiter).

% For the running environment,  we deploy {\color{red}\texttt{FATE} on a standalone docker image allocated with $20$ CPU cores and $256$Gib memory and simulate  the PPML training scenario in this standalone deployment.}

%\begin{itemize}
%\item  \texttt{FATE} provides many default implementations of machine learning algorithms, including logistic regression and tree-based algorithms. In our experiment, we use \texttt{FATE}'s default implementation of privacy-preserving logistic regression~\cite{yang2019acm}. This implementation employs a third-party arbiter to coordinate the process of gradient computation and send back the decrypted gradient to the participants. For data privacy protection, this implementation uses an RSA private key to encrypt the intermediate output which can only be decrypted by the public key owner (i.e., the arbiter).

%\item \texttt{PySyft} is a library for secure and private machine learning. Based on its communication and encrypted computation modules, we implement the VFL protocol of privacy-preserving logistic learning proposed in \cite{stephen2017corr}, which is similar to the \texttt{FATE}'s default implementation. Besides \cite{stephen2017corr}, there also exist many other similar learning protocols for privacy-preserving logistic regression~\cite{mohassel2017sp, yang2019acm}. We choose to implement the protocol provided in \cite{stephen2017corr} as it has good training accuracy as well as learning efficiency.

%\end{itemize}

\noindent\textbf{Attack setting.} We vertically partition each of the training datasets into two parts and distribute them to participants $\mathcal{A}$ and $\mathcal{B}$. Table~\ref{table:vertical_partition} summarizes the partition statistics. The labels of the three datasets are all placed in $\mathcal{A}$. In other words, $\mathcal{A}$ is the active participant while $\mathcal{B}$ is the passive participant.

For training and testing split,  we randomly split each of the three datasets into two parts, $80\%$ of the samples are taken as the training dataset while the other $20\%$ are taken as the testing dataset. Note that $\mathcal{A}$ and $\mathcal{B}$ has the same set of training IDs and the same set of testing IDs.

\begin{table}[!h]
\centering
\caption{Vertical partition of the datasets.}
\resizebox{0.45\textwidth}{!}{
\begin{tabular}{c|c|c|c} 
\hline \textbf{Dataset} & \textbf{\#samples} &\textbf{\#features for $\mathcal{A}$} & \textbf{\#features for $\mathcal{B}$} \\
\hline
\hline Credit & $30, 000 $ & $13$ & $10$ \\
 Breast & $699$ & $10$ & $20$ \\
 Vehicle & $ 946$ & $9$ & $9$ \\
\hline 
\end{tabular}}
\label{table:vertical_partition}
\end{table}

In the reverse multiplication attack, the activate participant is the adversary and therefore we set $\mathcal{A}$ to be the adversary. Here, the goal of $\mathcal{A}$ is to steal $\mathcal{B}$'s private data using limited information received from $\mathcal{B}$. During the training process, $\mathcal{A}$ stores the coefficients (or gradients) and the encrypted multiplication results. 
% In this experiment, the adversary only gathers information out of the training process, not destroying the training process. 

\subsection{Results and Analysis}
\label{reverse_sum_attack_exp}

\begin{figure}[!ht]
\centering
\includegraphics[width=0.5\textwidth]{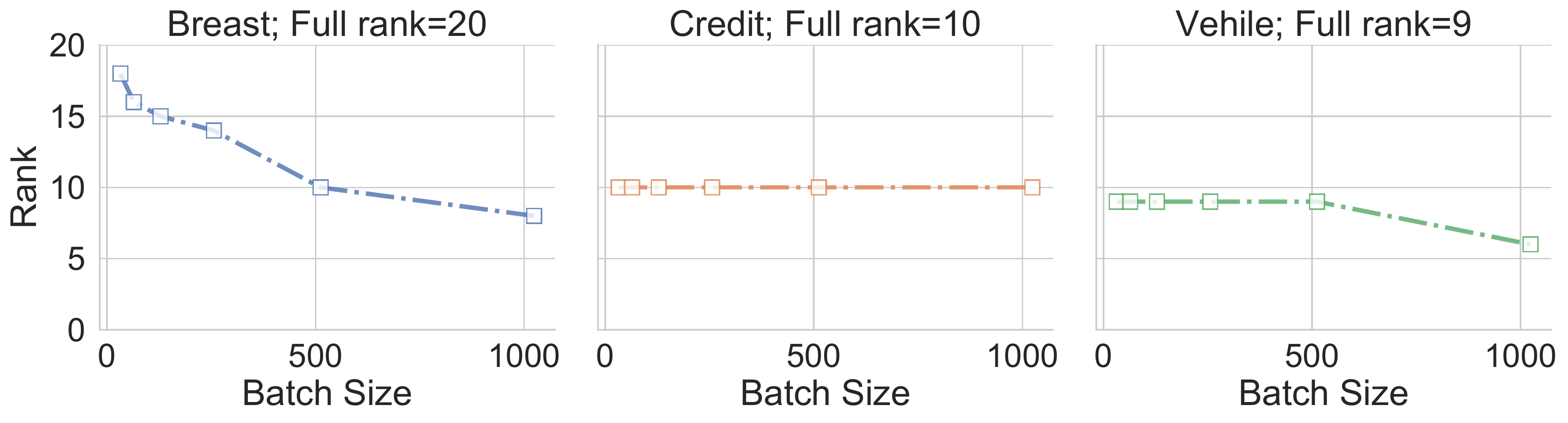}
\caption{The rank of coefficient matrix under different batch sizes.}
\label{fig:batch_size}
\end{figure}

\begin{figure}[!ht]
\centering
\includegraphics[width=0.5\textwidth]{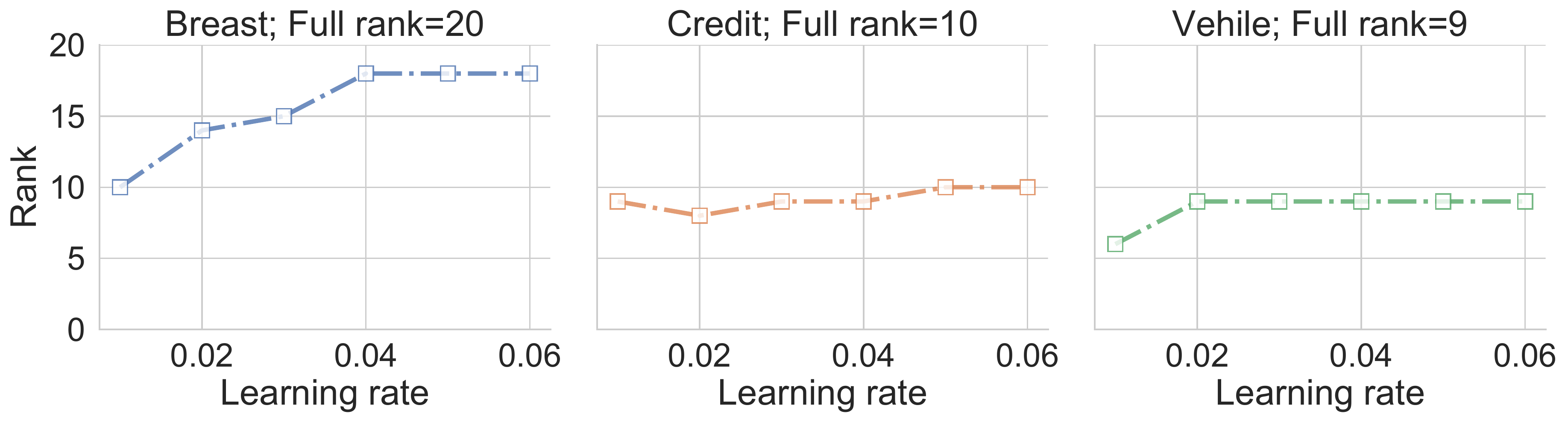}
\caption{The rank of coefficient matrix under different learning rates.}
\label{fig:learning_rate}
\end{figure}

% \noindent\textbf{Impact of training parameters.}  
As stated in \S\ref{sec:collusion_attack}, the success of the reverse multiplication attack depends on the rank of the coefficient matrix. The attack is successful only when the coefficient matrix is of full rank.

% which will be influenced by the training parameters. To understand the influence, we test our reverse multiplication attack under different settings of two critical training parameters: learning rate and batch size.

We first test the strength of our attack when different batch sizes are used for training.
Figure~\ref{fig:batch_size} reports the rank of the coefficient matrix after $100$ training epochs with different batch sizes. On each dataset, the testing accuracy with different batch size changes little almost remains the same. It shows that batch size can significantly affect the matrix rank, and small batch sizes can lead to larger ranks.
% This indicates that the adversary can steal almost all of the other participant's private data if an extremely small back size (e.g. 32, 64, 128) is used. 
The matrix rank is very stable on datasets Credit and Vehicle, and the coefficient matrix is of \emph{full rank} when the batch size is smaller than 500. This indicates that, on these two datasets, participant $\mathcal{A}$ can steal all of the private data from participant $\mathcal{B}$ using our attack. However, our attack fails on the Breast dataset. This indicates that some datasets are naturally more robust or vulnerable to our attack than others, causing more uncertainties in real-world scenarios.

We further test the strength of our attack when different learning rates are used for training. On each dataset, the testing accuracy with different learning rates changes little, less then $5\%$. Figure~\ref{fig:learning_rate} reports the rank of the coefficient matrix after $100$ training epochs with different learning rates.
One can see that learning rate demonstrates a quite opposite effect to batch size. Smaller learning rates tend to increase the matrix rank. Again, the rank is very stable on datasets Credit and Vehicle, and the full rank is obtained when the learning rate is larger than 0.02 on both datasets. In this case, our attack can successfully steal all of participant $\mathcal{B}$'s private data. Unfortunately, the attack still fails on the Breast dataset. We conjecture this is because Breast is a very sparse dataset which has many features but very few samples. This implies one limitation of our attack. However, it is worth noting that many real-world datasets are not sparse datasets, where our attack is still valid.

\section{Reverse Sum Attack}
\label{sec:reverse_sum_attack}

 %The sum is a common statistical factor, widely used in many machine learning algorithms, including nearest neighbor-based and tree-based algorithms~\cite{friedman2001elements}. 
 The reverse sum attack is designed for SecureBoost.  In this attack, the adversary attempts to reverse each addition term from the sum (see Eq.~\eqref{eq:reduction_loss}).

% Below, we describe our design of the reverse sum attack and then report our major experimental evaluation. 

\subsection{Attack Pipeline}
We assume two participants in SecureBoost including the activate participant $\mathcal{A}$ and the passive participant $\mathcal{B}$, and participant $\mathcal{A}$ is the adversary.  As introduced in $\S$\ref{sec:learning_protocol},  $\mathcal{A}$ knows: (1) the first and second gradient sums of the samples from $\mathcal{B}$'s data bins; and (2) the order of data bins.  
Thus, it is possible for $\mathcal{A}$ to first encode the information about the training data in the gradients, then reverse the addition terms of the gradient sums received from $\mathcal{B}$ using the encoded information.  These addition terms reveal to some extent the order of samples for $\mathcal{B}$'s feature.  For simplicity, we call such knowledge as \emph{partial order}, as defined below.

\vspace{0.05in}
\noindent \textbf{Partial Order.} Let $B_j^k$ denote the $k$-th data bin of the $j$-th feature.  For $\mathbf{X}$, the partial order of the $j$-th feature is formulated as $P_j = \left\{k|\forall \mathbf{x}_i \in \mathbf{X}, \exists B_j^k, \mathbf{x}_i \in B_j^k\right\}$.

\begin{figure}[!ht]
\centering
\includegraphics[width=0.48\textwidth]{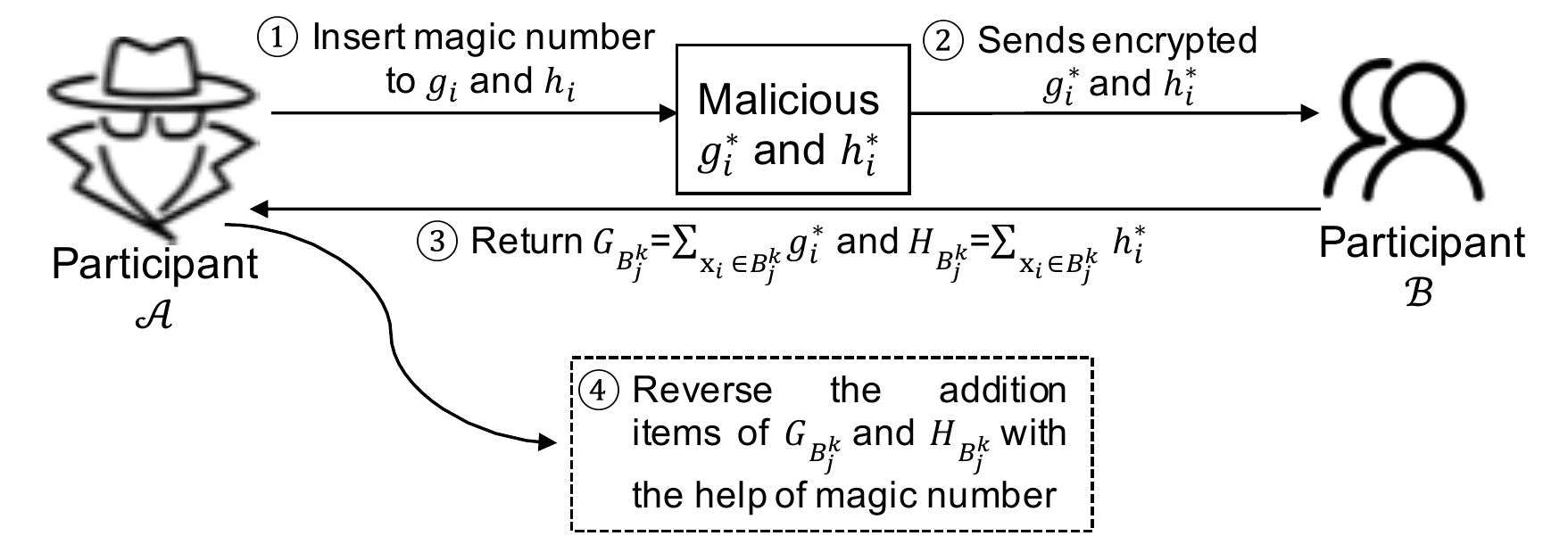}
\caption{The pipeline of the reverse sum attack.}
\label{fig:reverse_sum_attack_pipeline}
\end{figure}

%These addition terms imply the partial orders of the target participant's private training data.  
Figure~\ref{fig:reverse_sum_attack_pipeline} illustrates the pipeline of our proposed reverse sum attack, where the adversary aims to infer all the partial orders of $\mathcal{B}$'s private feature.
The major four steps of our attack are listed below.
% In our attack, we assume the adversary fully controls participant $\mathcal{A}$. The major two steps of our attack are listed below.
\begin{itemize}
\item{\textbf{Step 1.}} The adversarial participant $\mathcal{A}$ encodes a \textit{magic number}~\cite{magicnumber} into the first and second order gradients. Let $g_i$ and $h_i$ denote the clean gradients,  while $g_{i}^{*}$ and $h_{i}^{*}$ denote the magic number-encoded gradients. The magic number refers to distinctive unique values (i.e., global unique identifiers) that are unlikely to be mistaken for other meanings.

\item{\textbf{Step 2.}} The adversary encrypts and sends $g_{i}^{*}$ and $h_{i}^{*}$ to participant $\mathcal{B}$.

\item{\textbf{Step 3.}}  $\mathcal{B}$ calculates the first and the second gradient sums of each data bin, denoted as $G_{B_j^k}$ and $H_{B_j^k}$, respectively, and then sends them back to $\mathcal{A}$.

\item{\textbf{Step 4.}}  The adversary stores both the first and second gradient sums received from $\mathcal{B}$. It then reverse engineers all the addition terms from the gradient sums by leveraging the encoded magic numbers. These reversed terms will reveal the partial orders of $\mathcal{B}$'s features. 
\end{itemize}

We will provide more details of the above four attack steps in the following subsections.

% \vspace{0.05in}
\subsubsection{Magic Number} 
A magic number is a unique identifier used by the attacker to identify the specific gradient value encoded with this magic number.  We design a magic number generation schema that can generate as many magic numbers as possible to help identify the gradients of the training data. 

Specifically, we propose to encode magic numbers into the first and second gradients. The practical public encryption ecosystem encrypts on the $1024$-bit number. For the $64$-bit float number of the gradient, the ecosystem usually pads it with $960$ zeros to produce a $1024$-bit number for encryption. The $960$ zeros do not deteriorate the precision of the float number as they are padded at the least significant positions.  Instead of padding 960 zeros, in our attack, we propose to pad the $64$-bit number with a $960$-bit magic number.
% such ciphertext is stored as large numbers of integers, occupying $1020$ bits in computer memory.
% the numeric value of the first and second order gradient is stored as double-precision floating point number, occupying 64 bits in computer memory. 
% In this format, the last $52$ bits determines the significant precision of the number, giving from $15$ to $17$ significant decimal digits precision ($2^{-53} \approx 1.11 \times 10^{-16}$)~\cite{float64}. 
\begin{figure}[!ht]
\centering
\includegraphics[width=0.5\textwidth]{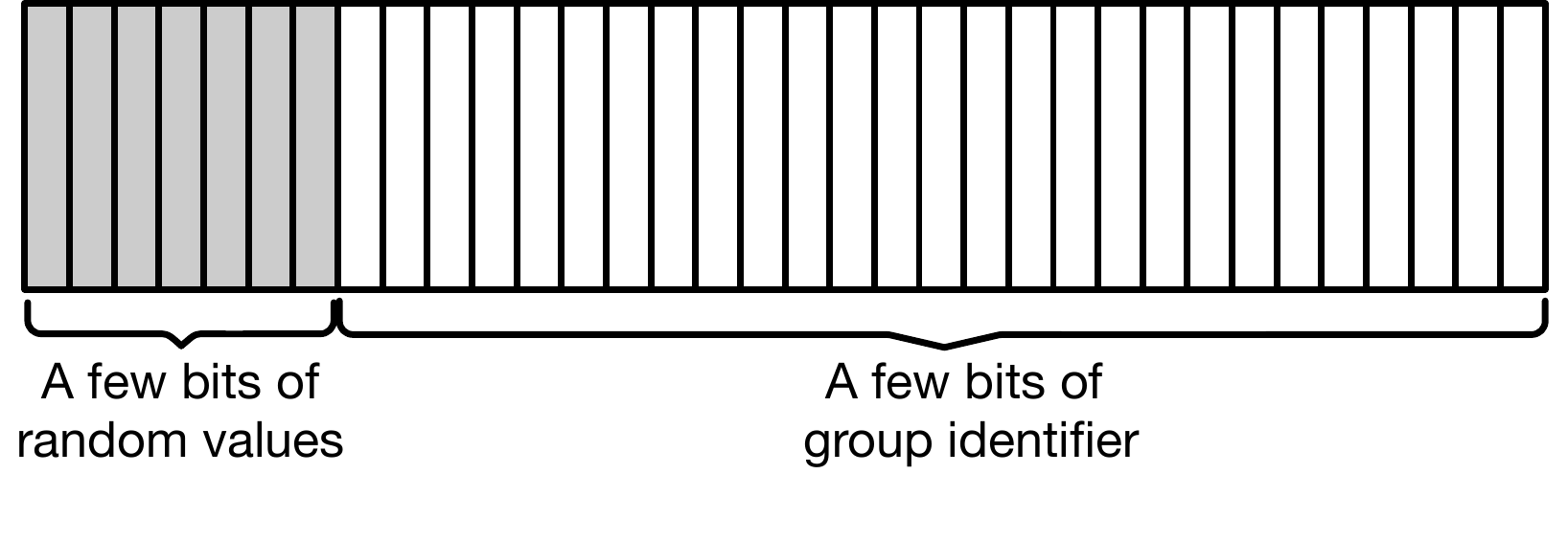}
\caption{The architecture of the magic number. 
The first few bits are random values while the remaining bits are  group identifier that allows the adversary to locate the specific group the sample belongs.}
\label{fig:magic_number}
\end{figure}

Figure~\ref{fig:magic_number} illustrates the architecture of our magic number schema. The magic number has two parts: \textit{group identifier} and \textit{random values}. The group identifier is a one-hot vector under a $n$-base positional numerical system and will be used to identify the specific group the encoded samples belongs to. For example, in a base-10 system, \texttt{0$\cdots$01} indicates that the sample belongs to group $1$.  Random values are used to differentiate samples from the same group. With the group identifier and random values, the  samples from the same group have the same group identifier while different random values.

\noindent \textbf{An example of magic number.} As an example, here we show how to generate the group identifiers using a base-$16$ positional numerical system.  For simplicity, we only utilize 20 out of the 960 bits to encode the magic number. We set the first 4 bits as random value bits and set the remaining $16$ bits as the group identifier bits, which can identify a maximum of 4 ($= \log_{16} {{2}^{16}}$) unique groups of samples. Given $60$ data samples, we first divide them into $4$ groups with each has $15=60/4$ samples.  The $1$-st to $15$-th samples are placed in the first group, the $16$-th to $30$-th samples are in the second group, the $31$-st to $45$-th samples are in the third group, and the $46$-th to $60$-th samples are in the fourth group. For samples in the first group, we generate random values for the first 4 bits and set the group identifier in the remaining $16$ bits to \texttt{0X0001}. Similarly, for samples in the second group, the first 4 bits of the magic number are random values and the group identifier is \texttt{0X0010}.

\begin{algorithm}[!ht]
\caption{Gradient Encoding}
\label{alg:lsb_encoding}
\begin{algorithmic}[1]
\State \textbf{Input:} the first order gradients $g_1, g_2, \dots, g_n$, the second order gradients $h_1, h_2, \dots, h_n$. % and the previous tree $\mathcal{T}.$
\State \textbf{Output:} the first and second gradients encoded with magic numbers.
\State $k \gets$ number of super groups 
\State $b \gets$ base of the positional numerical system
\State $l \gets \log_b 2^{(960 - 30 \times k)}$ \Comment{length of the group identifier}
\State $g \gets 2 \times k \times l$ \Comment{number of the total groups}
\State $n' \gets g \times b$  \Comment{number of the encoded samples}
\For{$i=1;i<=n'; i++$}
	\State $s \gets$ round($i / l$)
	\State $id \gets$  round($(i \% l )/(b-1)$)
	\State $random~values \gets$ the $s$ to $s+29$ bits are set to random values and the remaining of the first $30\times k$ bits are set to $0$
	\State $group~identifier \gets $ the one-hot vector under $b$-base positional numeral system with the $id$-th number set to $1$
	\State $magic~number \gets $ concatenate($random~values$, $group~identifier$) 
%	\State $C_g \gets \{g_i|\forall \mathbf{x}_i,  \mathbf{x}_i \in \ell\}$ 
%	\State $C_h \gets \{h_i|\forall \mathbf{x}_i,  \mathbf{x}_i \in \ell\}$
%	\State $G \gets G \cup \text{Random(}C_g, \frac{s}{v}$) \Comment{Select gradients}
%	\State $H \gets H \cup \text{Random(}C_h, \frac{s}{v}$)  \Comment{Select hessian matrixes}
	\EndFor

%\State $G^* \gets \{\text{MagicNumber}(g_i) |g_i \in G\}$  \Comment{Figure~\ref{fig:magic_number} shows encoding}
%\State $H^* \gets \{\text{MagicNumber}(h_i) |h_i \in H\}$
\end{algorithmic}
\end{algorithm}

\subsubsection{Gradient Encoding}
%Algorithm~\ref{alg:lsb_encoding} shows how to encode magic numbers into the least significant bits of the first and second gradients. 

%Algorithm~\ref{alg:lsb_encoding} shows how to encode magic numbers into the least significant bits of the first and second gradients. 

Algorithm~\ref{alg:lsb_encoding} describes how to encode the magic numbers into the gradients. 

\textbf{First}, we set the number of supergroups to $k$ and the positional numerical system base to $b$.  In this setting, the  maximum length of the group identifier is $\log_b 2^{(960 - 30 \times k)}$, shown in Line~5. The maximum number of sample groups is $2\times k \times l$ and the maximum number of encoded samples is $g\times b$, as shown in Lines~6\&7 respectively. \textbf{Second}, we sequentially select $n'$ samples to encode magic numbers into their first and second gradients. Other selection strategies like distribution-based selection can also be used here. \textbf{Third}, for each selected sample, we construct a magic number, then post process its first or second gradients by setting the lower $b$ bits to the bit string of the magic number. Lines 9--13 describe the detailed steps of constructing a magic number while the architecture of the magic number is shown in Figure~\ref{fig:magic_number}.

% Third, for each selected samples, we construct a magic number and post-process its first or second gradients by setting the lower $b$ bits of each to a bit string of the magic number. Lines 9--13 describe the detailed steps of constructing a magic number, and Figure~\ref{fig:magic_number} shows the architecture of the magic number.
%select a set of the first order gradients and a set of the second order gradients according to a certain strategy.  Lines 7--11 introduce a strategy for selecting gradients based on the distribution of leaf nodes. We can also use a simple random selection strategy.
% Second, post process the first and second gradients by setting the lower $b$ bits of each to a bit string $s$ of magic number. 
%Figure~\ref{fig:magic_number} shows the design of magic number. In our design, the first $7$ bits are set to random values and the remaining $27$ bits are set to a certain bit string. Below, we detail our design of the magic number.

\subsubsection{Gradient Sum Reversion} By leveraging the encoded magic number, the adversary now can effectively reverse engineer all the addition terms from the gradient sum.

%Let $B$ denote the bin. Then, let $G_{B}=\sum\nolimits_{\mathbf{x}_i \in B} g_i$ and $H_{B}=\sum\nolimits_{\mathbf{x}_i \in B } h_i$ be the first and second order gradient sum of $B$'s samples. 
During the collaborative training of each boost tree, the adversary $\mathcal{A}$ will continually receive $G_{B_j^k}$ and $H_{B_j^k}$ from participant $\mathcal{B}$. The adversary's goal is to reverse all the addition items of $G_{B_j^k}$ and $H_{B_j^k}$, all of which can be further integrated to infer the partial orders.

Given $G_{B_j^k}$ and $H_{B_j^k}$,  we first decode it into $1024$-bit strings, then extract the lower $960$ bits (denoted as $s$). In each $G_{B_j^k}$ or $H_{B_j^k}$, $s$ equals to the sum of the magic numbers of $B_j^k$'s samples encoded by the adversary. Then, by leveraging $s$, the adversary greedily searches an optimal combination of the samples that might be the addition terms of $G_{B_j^k}$ or $H_{B_j^k}$. Next, we will use an example to illustrate how the adversary reverse engineers all the addition items by leveraging $s$.

In this example, we utilize $20$ of the $960$ bits to encode the magic number, take the first $4$ bits as random value bits, and set the base of the positional numerical system to $4$. We assume a data bin $B_j^k$ containing five samples:  $\mathbf{x}_{2}$,  $\mathbf{x}_{17}$,  $\mathbf{x}_{517}$, $\mathbf{x}_{520}$, $\mathbf{x}_{2400}$,  where only $\mathbf{x}_{2}$ and $\mathbf{x}_{17}$ are selected to encode the magic numbers. The magic number of  $\mathbf{x}_{2}$ is \texttt{0X30001} and the magic number of $\mathbf{x}_{14}$ is \texttt{0X20010}.
In this example, the magic number sum of $G_{B_j^k}$ is \texttt{0X50011} =  \texttt{0X30001} + \texttt{0X20010}. Once the magic sum of \texttt{0X50011} is extracted, the adversary can know that $B_j^k$ must contain one sample from group $1$ and one sample from group $2$.  With the additional information, i.e., the sum of random values of the two magic numbers is  \texttt{0X5}, the adversary can now easily figure out that $\mathbf{x}_{2}$ and $\mathbf{x}_{17}$ are in $B_j^k$.
 
% as an example, we assume only $\mathbf{x}_{14}$ and $\mathbf{x}_{1400}$ are selected to encode magic numbers.
% into the first and second order gradients. 
%For $\mathbf{x}_{14}$, the magic number of  $g_{14}$ is \texttt{0b0111000000001000000000000000000}  and the magic number of $h_{14}$ is \texttt{0b1000001000000000000000000000000000}.

% For $\mathbf{x}_{1400}$, the magic number  of $g_{2400}$ is \texttt{0b10110010000000000000000\\00000000000} and the magic number of $h_{2400}$ is \texttt{0b10110010000000\\000000001000000000}.
 % In this example, the adversary extracts the magic number sum of \texttt{0b1110000000000001000000000000000000} from $G_{K}$, and the magic number sum of \texttt{0b00110100000000000000\\00001000000000} from $H_{K}$.
% \texttt{0b11100000000000010000000000000\\00000} clearly indicts $\mathbf{x}_{14}$ must be in the bucket, and \texttt{0b00110100000\\00000000000001000000000}  indicts  $\mathbf{x}_{2400}$ must be in the bucket.

\subsection{Experimental Setup}

\noindent\textbf{Datasets.} 
We conduct experiments on two public datasets: Credit and Student~\cite{student}.  The Credit dataset is the same to the one used in $\S\ref{sec:experiment_multiplication}$. The Student dataset is a regression dataset, containing education records of $395$ students collected from a mathematics class.  The data attributes include student grades, demographic, social, and school-related features. This dataset is often used to train a regression model that predicts student performance in the final examination.  

\vspace{0.05in}
\noindent\textbf{Setup for SecureBoost.} 
We perform the reverse sum attack against the real-world implementation of SecureBoost in \texttt{FATE}. Similar to $\S~\ref{sec:experiment_multiplication}$, we use \texttt{FATE}'s default implementation of the SecureBoost learning protocol. This implementation utilizes Paillier to protect data privacy by encrypting the first and second order gradients, and the intermediate outputs.

\vspace{0.05in}
\noindent\textbf{Attack setting.} 
We vertically partition each of the training datasets into two parts and distribute them to participants $\mathcal{A}$ and $\mathcal{B}$. The training and testing split is the same as $\S$ \ref{sec:experiment_multiplication}. Table \ref{table:vertical_partition_sum} summarizes the partition statistics. Participant $\mathcal{A}$ is the activate participant who holds the class labels, while participant $\mathcal{B}$ is the passive participant. 

\begin{table}[!h]
\centering
\caption{Vertical partition of the datasets.}
\resizebox{0.45\textwidth}{!}{
\begin{tabular}{c|c|c|c} 
\hline \textbf{Dataset} & \textbf{\#samples} &\textbf{\#features for $\mathcal{A}$} & \textbf{\#features for $\mathcal{B}$} \\
\hline
\hline Credit & $30, 000 $ & $13$ & $10$ \\
 Student & $395$ & $6$ & $7$ \\
\hline 
\end{tabular}}
\label{table:vertical_partition_sum}
\end{table}

In the reverse sum attack, the activate participant is the adversary. We set the activate participant $\mathcal{A}$ to be the adversary who aims to steal the partial orders of $\mathcal{B}$'s private training data. During training, the adversary encodes magic numbers into the least significant bits of the gradients and stores the gradient sums from participant $\mathcal{B}$. The adversary only slightly modifies the gradients at the least significant bits, thus does not cause much negative impact to the training process (the selection of the best split dimension and feature).
% since the least significant bits cannot affect the selection of the best split dimension and feature. 

\subsection{Results \& Analysis}
%\begin{figure}[!ht]
%\centering
%\includegraphics[width=0.45\textwidth]{secureboost_impact_encoding_setting}
%\caption{The number of successfully stolen examples.}
%\label{fig:secureboost_encoding_setting}
%\end{figure}
%
%\begin{figure}[!ht]
%\centering
%\includegraphics[width=0.45\textwidth]{fig/secureboost_success_rate_encoding_setting}
%\caption{The percentage of successfully stolen examples in  all  the examples with encoded magic numbers.}
%\label{fig:secureboost_success_rate}
%\end{figure}

Here, we test the performance of our reverse sum attack under different data distributions, bin sizes and attack parameters. We measure the attack success rate by the percentage of exactly reverse engineered partial orders among all the samples inserted with magic number. We also evaluate the value of the leaked data by training an alternative classifier on the leaked data.
% In this experiment, we investigate the performance of the reverse sum attack against SecureBoost in vertical federated learning. Specifically, we experimentally analyze the factors that influence the performance of the attack, including the distribution of the data, the bin size, and the attack parameters.

\begin{table}[!h]
\centering
\caption{Details of the four $1$D synthetic datasets.}
%\resizebox{0.45\textwidth}{!}{
\begin{tabular}{c|c|c} 
\hline \textbf{Dataset} & \textbf{\#samples} & \textbf{distribution} \\
\hline
\hline $D_1$ & $30, 000$ & Normal distribution: $\mathcal{N}(0, 1)$  \\
 $D_2$ & $30, 000$  & Bernoulli distribution: $\mathcal{B}(\frac{1}{2})$  \\
 $D_3$ &$30, 000$  & Exponential distribution: $\mathcal{E}(1)$\\
 $D_4$ &$30, 000$  & Uniform distribution:  $\mathcal{U}(0, 50)$\\
\hline 
\end{tabular}
%}
\label{table:synthetic_dataset}
\end{table}

\begin{figure}[!ht]
\centering
\includegraphics[width=0.45\textwidth]{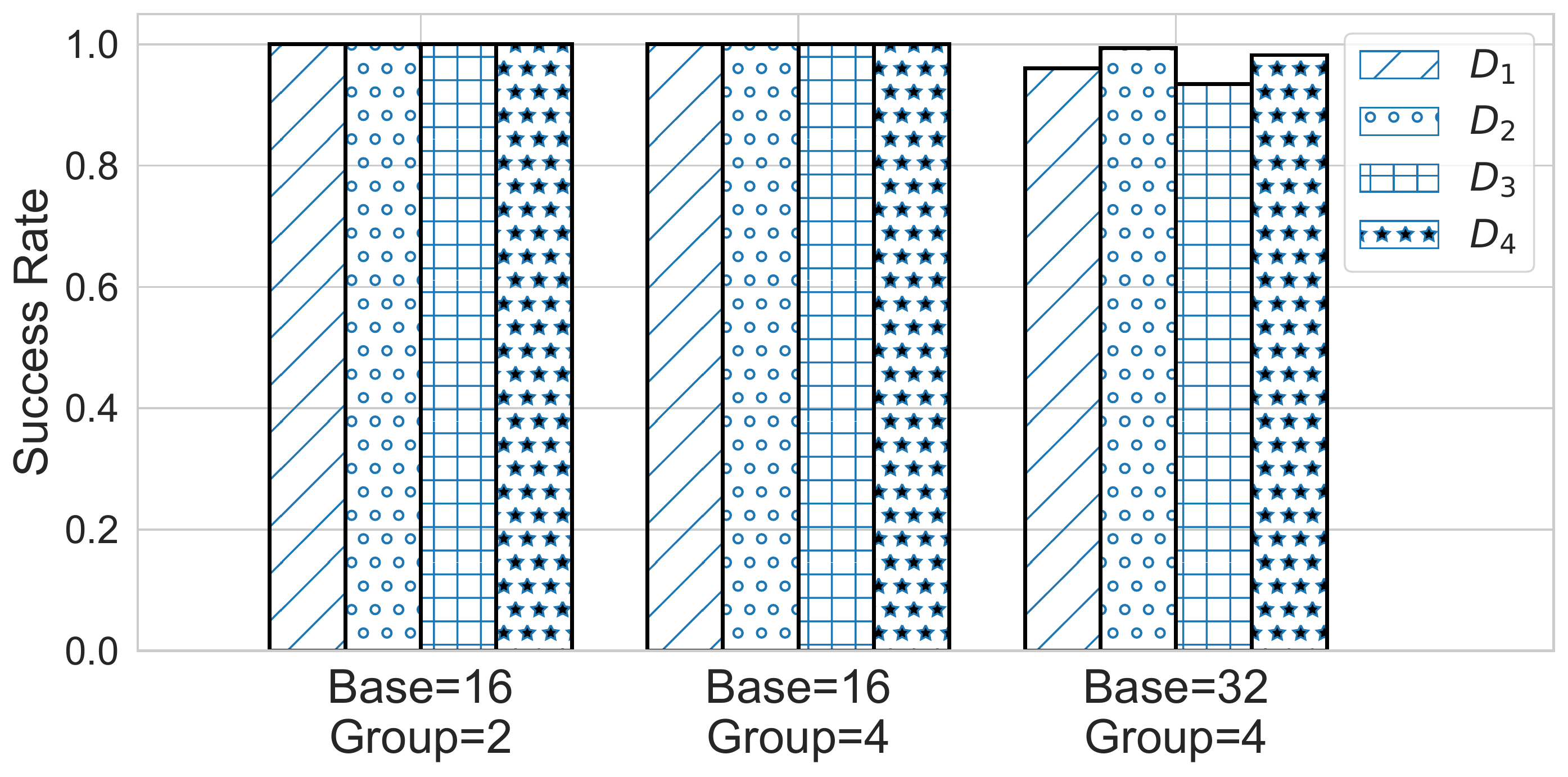}
\caption{Attack success rate on four synthetic datasets using different attack parameters (Base: $b$; Group: $k$).}
\label{fig:secureboost_data_distribution}
\end{figure}

\vspace{0.05in}
\noindent\textbf{Performance under different data distributions.} We first perform our attack independently on four synthesized $1$D datasets of different data distributions, including  Normal distribution, Bernoulli distribution, Exponential distribution, and uniform distribution.  The statistics of the four synthetic datasets is summarized in Table~\ref{table:synthetic_dataset}.
Figure~\ref{fig:secureboost_data_distribution} shows the attack success rate on the four synthetic datasets under different settings of attack parameters including the base ($b$) of the positional numerical system and the number ($k$) of supergroups.  As can be observed, our attack can accurately (success rate $>95\%$) reverse engineer the partial orders of participant $\mathcal{B}$'s private data, consistently across all four synthetic datasets and different attack parameters. 

%We conjecture that 

\begin{figure}[!ht]
\centering
\includegraphics[width=0.5\textwidth]{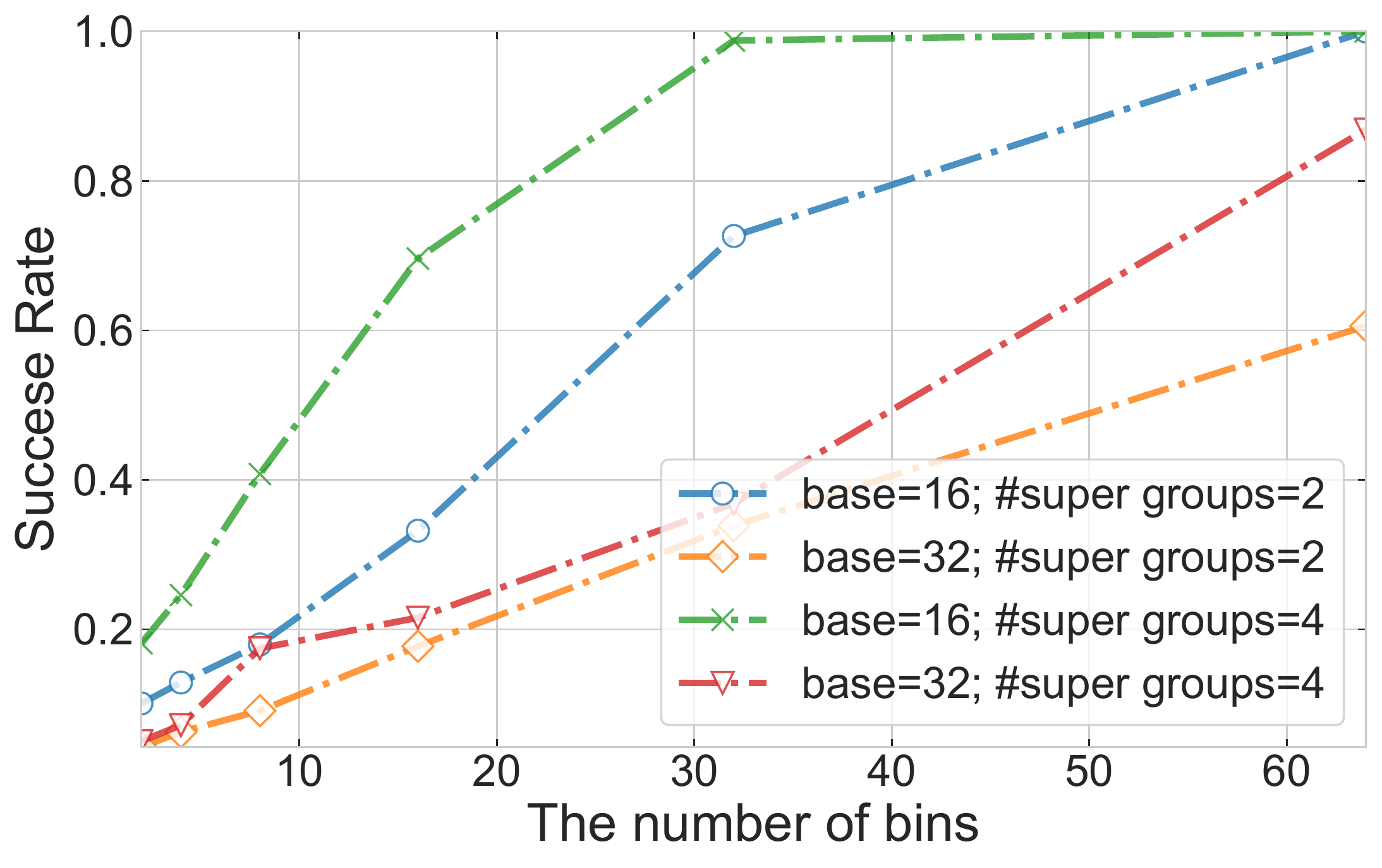}
\caption{Attack success rate on synthetic dataset $D_1$ under different number of data bins.}
\label{fig:secureboost_data_bins}
\end{figure}

\vspace{0.05in}
\noindent \textbf{Performance under different bin sizes.} 
Here, we explore the impact of bin size on the attack success rate. We set two participants to jointly learn an XGBoost model on the Student dataset. % $D_1$ 
Figure~\ref{fig:secureboost_data_bins} shows the attack success rate when the target participant $\mathcal{B}$ splits its training data into different numbers of bins.
As can be observed, the attack success rate increases with the increase of the number of data bins used by the target participant. This is because more data bins tend to result in fewer samples in a single bin, making it easier for the adversary to accurately figure out all the addition terms of the gradient sum of this bin. The attack success rate is above $50\%$ even when an extremely small number (close to $30$) of data bins are used.

\begin{table*}[!h]
\centering
\caption{Attack success rate under different attack parameters. FID: feature ID. }
%\resizebox{0.45\textwidth}{!}{
\begin{tabular}{c|c|c|c|c|c|c|c} 
\hline \multirow{2}{*}{\textbf{Dataset}} &  \multirow{2}{*}{\textbf{\#Super groups}} & \multirow{2}{*}{\textbf{FID}} & \multicolumn{5}{c}{\textbf{Base of the positional numerical system}} \\
\cline{4-8}
& & & 2 & 4 & 8 & 16 & 32 \\
\hline
\hline 
\multirow{20}{*}{Credit} & \multirow{10}{*}{2} &0& $1,860$ ($100.00\%$) & $2,784$ ($100.00\%$)  & $2,138$ ($49.27\%$) &  $2,379$ ($34.19\%$) & $202,8$ ($17.59\%$) \\
&& 1 & $1,860$ ($100.00\%$) & $1,047$ ($37.61\%$) & $876$ ($20.19\%$)  & $1,340$ ($19.26\%$) & $821$ ($7.12\%$) \\
&& 2 & $1,860$ ($100.00\%$)  & $1,218$ ($43.75\%$) & $1,203$ ($27.73\%$) & $1,150$ ($16.52\%$) & $639$ ($5.54\%$) \\
&& 3 & $1,860$ ($100.00\%$) & $1,070$ ($38.43\%$)  & $907$ ($20.89\%$) & $885$ ($12.72\%$) & $591$ ($5.12\%$) \\
&& 4 & $1,860$ ($100.00\%$) & $2,509$ ($90.11\%$) & $2,208$ ($50.87\%$) & $2,246$ ($32.27\%$) & $2,166$ ($18.78\%$) \\
&& 5 & $1,860$ ($100.00\%$)  & $1,470$ ($52.80\%$) & $1,121$ ($25.82\%)$ & $1,509$ ($21.68\%$) & $799$ ($6.93\%$) \\
&& 6 & $1,860$ ($100.00\%$) & $1,442$ ($51.78\%$)  & $877$ ($20.21\%$) & $1,136$ ($16.32\%$) & $1,113$ ($9.65\%$) \\
&& 7 & $1,860$ ($100.00\%$) & $1,463$ ($52.56\%$) & $915$ ($21.08\%$) & $1,034$ ($14.86\%$) & $680$ ($5.9\%$) \\
&& 8 & $1,860$ ($100.00\%$) & $1,443$ ($51.83\%$) & $1,100$ ($25.35\%$) & $1,394$ ($20.03\%$) & $814$ ($7.06\%$) \\
&& 9 & $1,860$ ($100.00\%$) & $1,458$ ($52.37\%$) & $1,223$ ($28.19\%$) & $1,370$ ($19.68\%$) & $853$ ($7.4\%$) \\
\cline{2-8}
& \multirow{10}{*}{4} &0 & $1,888$ ($100.00\%$) & $2,688$ ($100.00\%$) & $2,283$ ($54.35\%$) & $2,909$ ($43.29\%$) & $2,856$ ($25.59\%$) \\
&& 1 & $1,888$ ($100.00\%$) & $1,389$ ($51.66\%$) & $1,328$ ($31.62\%$) & $1,443$ ($21.48\%$) & $1,063$ ($9.52\%$) \\
&& 2 & $1,888$ ($100.00\%$) & $1,405$ ($52.26\%$) &  $1,366$ ($32.52\%$) & $1,494$ ($22.23\%$) & $1,071$ ($9.60\%$) \\
&& 3 & $1,888$ ($100.00\%$) & $1,406$ ($52.28\%$) & $1,174$ ($27.96\%$) & $1,326$ ($19.74\%$) & $802$ ($7.18\%$) \\
&& 4 & $1,888$ ($100.00\%$) & $2,688$ ($100.00 \%$) & $2,236$ ($53.25\%$) & $3,156$ ($46.97\%$) & $2,958$ ($26.51\%$) \\
&& 5 & $1,888$ ($100.00\%$) & $1,496$ ($55.67\%$) & $1,450$ ($34.52\%$) & $1,452$ ($21.6\%$) & $1,522$ ($13.64\%$) \\
&& 6 & $1,888$ ($100.00\%$) & $1,455$ ($54.13\%$) & $1,378$ ($32.8\%$) & $1,444$ ($21.49\%$) & $1,593$ ($14.27\%$) \\
&& 7 & $1,888$ ($100.00\%$) & $1,459$ ($54.28\%$) & $1,351$ ($32.17\%$) & $1,363$ ($20.29\%$) & $1,296$ ($11.61\%$) \\
&& 8 & $1,888$ ($100.00\%$) & $1,437$ ($53.47\%$) & $1,423$ ($33.89\%$) & $1,349$ ($20.07\%$) & $1,260$ ($11.29\%$) \\
&& 9 & $1,888$ ($100.00\%$) & $1,467$ ($54.59\%$) & $1,413$ ($33.64\%$)  & $1,537$ ($22.87\%$) & $1,553$ ($13.92\%$) \\
\hline
\end{tabular}
%}
\label{table:attack_parameters}
\end{table*}

\begin{figure*}[!ht]
\centering
\begin{minipage}[c]{0.48\textwidth}
\centering
\subfloat[\#super groups=2]{\includegraphics[width=0.50\textwidth, height=0.3\textwidth]{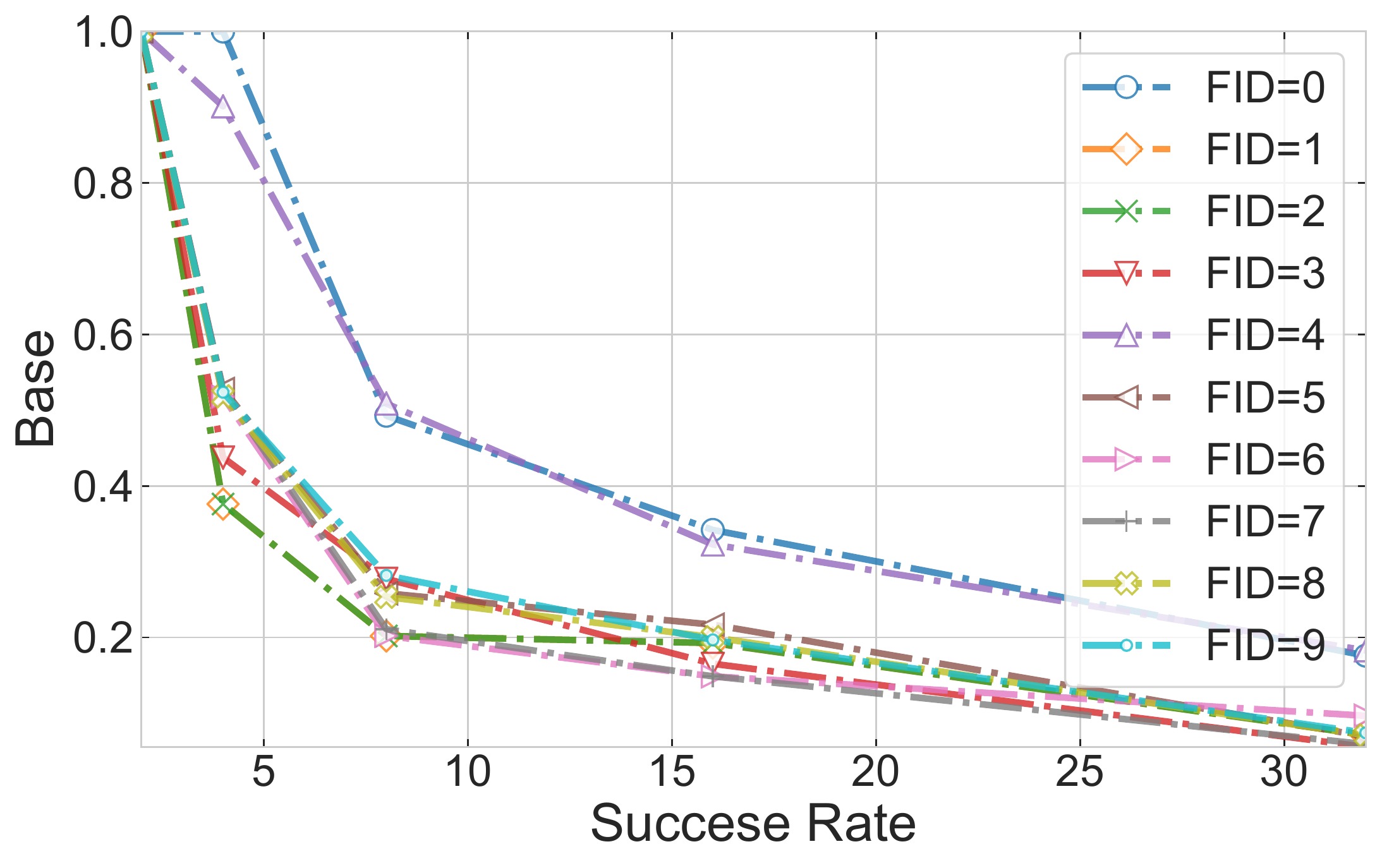}}
\subfloat[\#super groups=4]{\includegraphics[width=0.50\textwidth, height=0.3\textwidth]{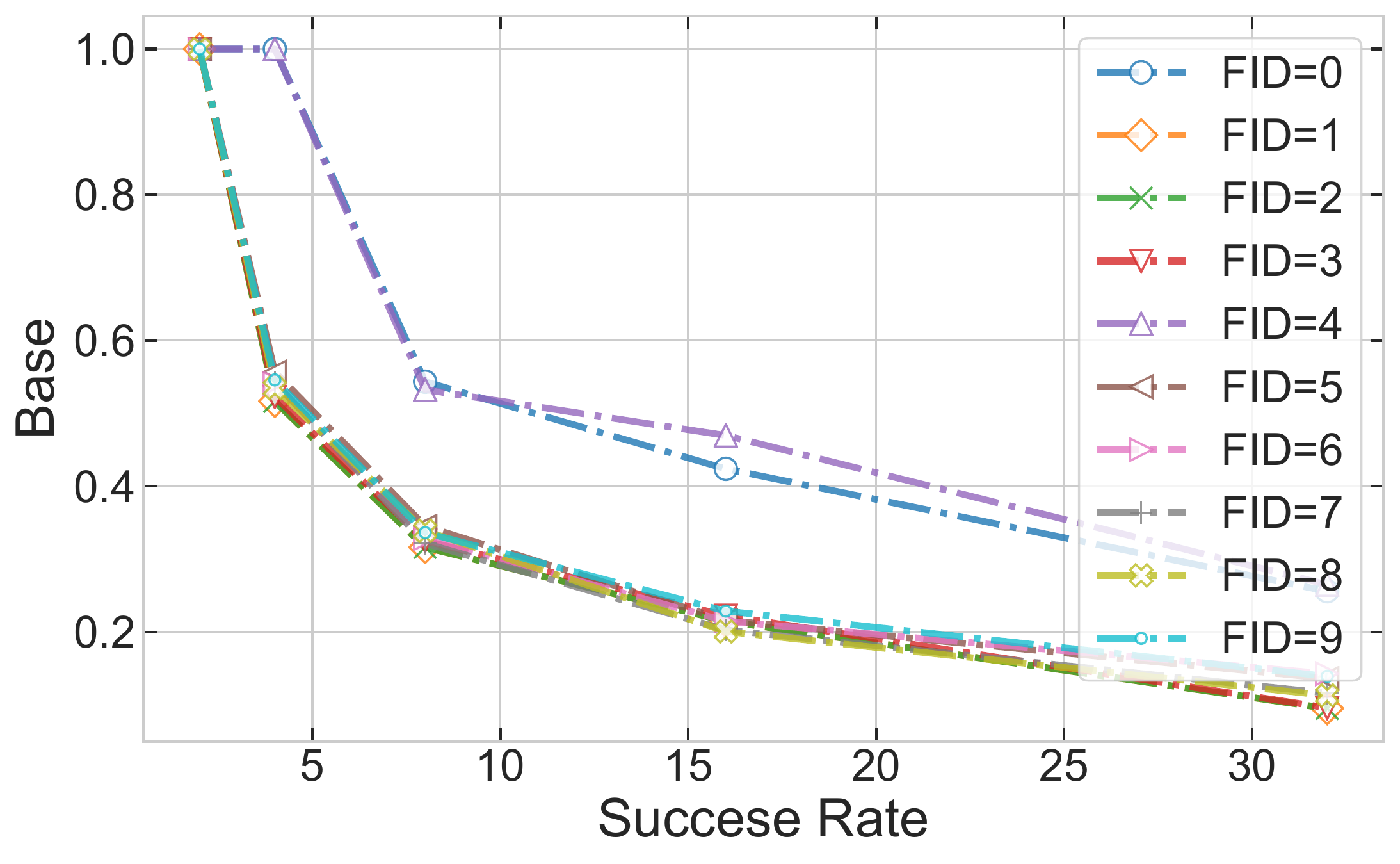}}
\caption{Attack success rate on the Credit dataset with different number of data bins and numerical bases.}
\label{fig:attack_success_rate}
\end{minipage}
\quad 
\begin{minipage}[c]{0.48\textwidth}
\subfloat[\#super groups=2]{\includegraphics[width=0.5\textwidth, height=0.3\textwidth]{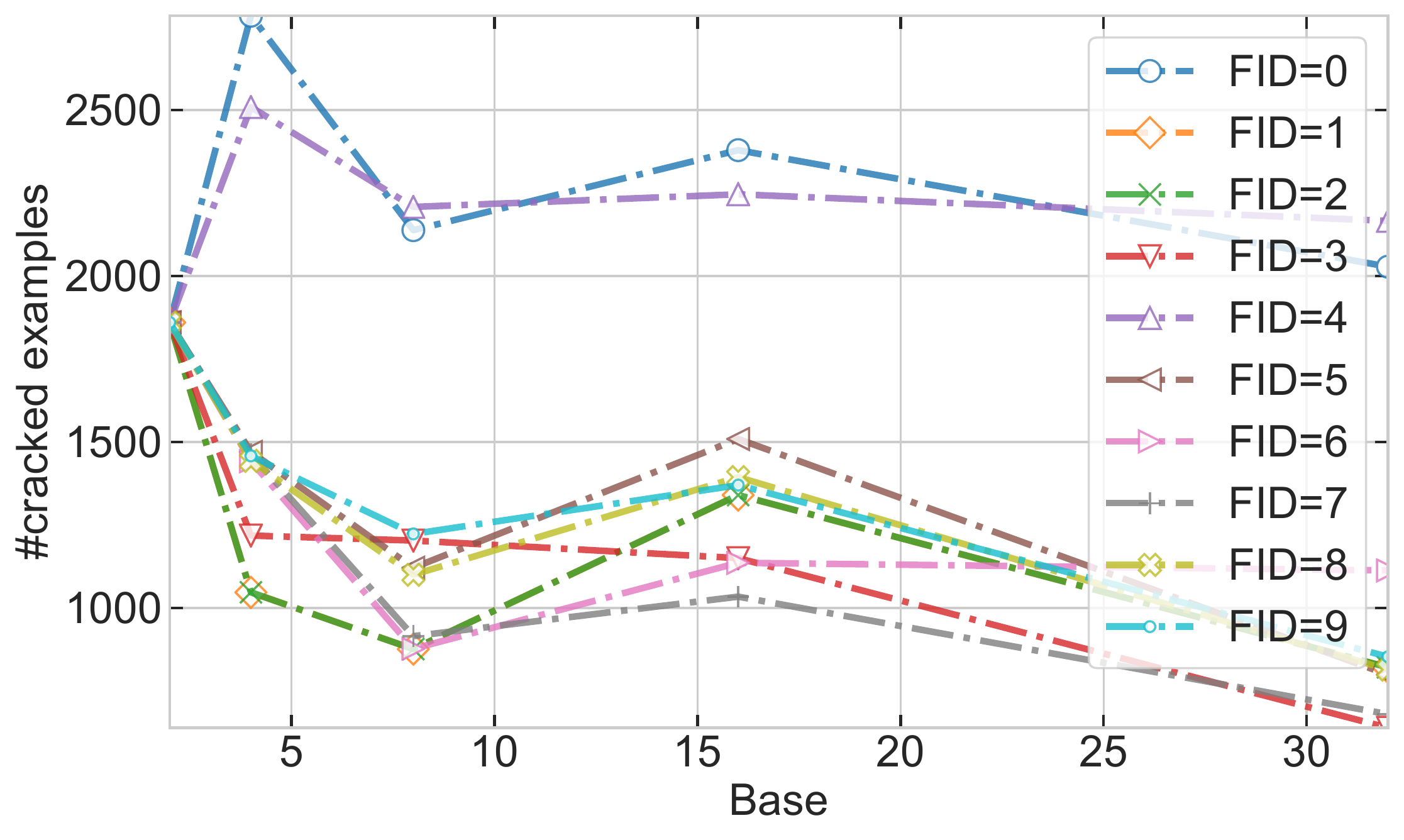}} 
\subfloat[\#super groups=4]{\includegraphics[width=0.5\textwidth, height=0.3\textwidth]{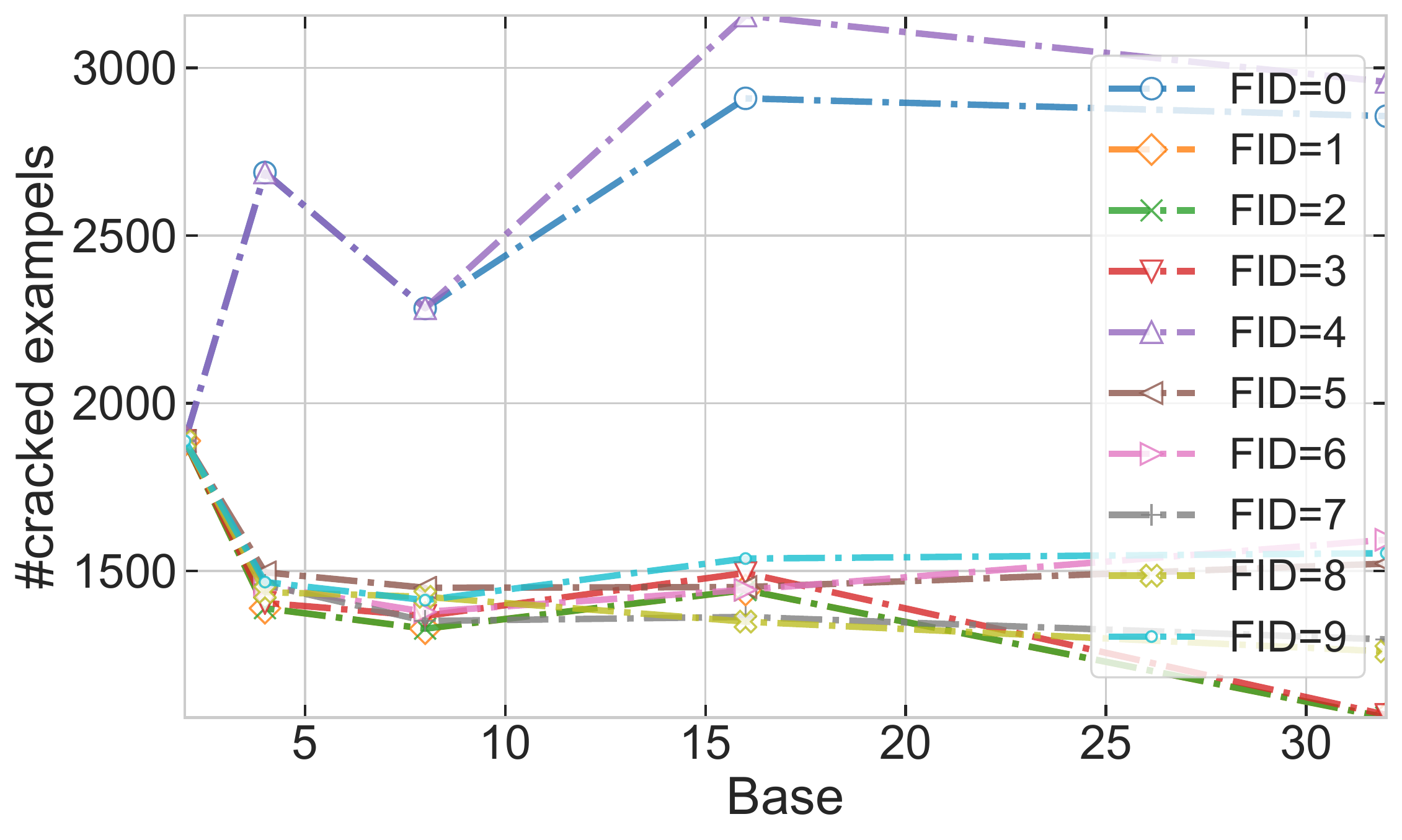}}
\caption{Number of cracked samples on the Credit dataset with different number of data bins and numerical bases.}
\label{fig:attack_cracked_number}
\end{minipage}
\end{figure*}

%\begin{figure}[!ht]
%\centering
%\subfloat[\#super groups=2]{\includegraphics[width=0.35\textwidth]{fig/secureboost_impact_sum_base_supergroup2.eps}} \\
%\subfloat[\#super groups=4]{\includegraphics[width=0.35\textwidth]{fig/secureboost_impact_sum_base_supergroup4.eps}}
%\caption{The success rate for the Credit dataset with different number of data bins and numerical base.}
%\label{fig:attack_cracked_number}
%\end{figure}
\vspace{0.05in}
\noindent \textbf{Performance under different attack parameters.} Here, we provide a more comprehensive analysis on the performance of our attack under the two attack parameters: the base ($b$) of the positional numerical systems and the number ($k$) of supergroups.
These two parameters are highly correlated with the number of partial orders needed to be cracked, as they decide the maximum number of samples that can be encoded with the magic numbers. In this experiment, we run the attack on the Credit dataset.

Table~\ref{table:attack_parameters} reports the success rate as well as the number of successfully cracked samples by our attack using different numerical bases and supergroup sizes. The results show that, using a small base and group size results in the highest attack success rate, while using a large base and a large group size results in more cracked samples. The reason behind this is twofold. By using a large base and a supergroup size, 
the capacity of the encoded samples ramps up. This basically requires the adversary to encode magic numbers into more samples, thus it can obtain more information about the addition terms of the gradient sum. However, the large capacity also creates a large search space for the samples in the same group, making it harder for the adversary to figure out all the right addition terms.
 
We further plot the attack success rates under different attack parameters in Figure~\ref{fig:attack_success_rate} and the number of cracked samples in Figure~\ref{fig:attack_cracked_number}.
As shown in Figure~\ref{fig:attack_cracked_number}, SecureBoost leaks more partial order information when the base of the numerical system and the number of supergroups increase, i.e., more cracked samples. 
%Figure~\ref{fig:secureboost_success_rate} shows the percentage of successfully stolen examples in all the examples that are encoded with magic numbers.
Comparing Figure~\ref{fig:attack_cracked_number} with Figure~\ref{fig:attack_success_rate}, we find that, even though a large base or a large supergroup size
makes SecureBoost leak more information, they  reduce the effectiveness of a single magic number as the success rate decreases.

% Figure~\ref{fig:secureboost_encoding_setting} plots the size of examples whose partial orders are successfully stolen by the adversary. 
% As we can see, SecureBoost leaks more partial order information when the  length of encoding base or the number of random value groups increases.

%\paragraph{Potential Side-effects}  
\subsection{Further Evaluation}
%\noindent\textbf{Value of the leaked information.}
Further, we measure the value of the leaked information by training an alternative classifier on the leaked information. To evaluate the alternative classifier, we propose to use a small auxiliary dataset, which follows the same distribution as $\mathcal{B}$'s features, to learn the mapping from the target participant's raw feature into the bin value (left and right bounds of the bin). In the following, we first introduce and evaluate the proposed mapping approach, then train an alternative classifier on the leaked information.

\begin{table*}[!h]
\centering
\caption{The percentage of the successfully inferred data bin values on the Credit dataset. FID denotes the feature ID of the target participant's training data; and $\#$bins denotes the number of data bins on the given feature. }
%\resizebox{0.45\textwidth}{!}{
\begin{tabular}{c|c|c|c|c|c|c|c} 
\hline \multirow{2}{*}{\textbf{FID/\#bins}} & \multicolumn{7}{c}{\textbf{\#samples known to the adversary}} \\
\cline{2-8}  & $5$ & $10$ & $20$ & $40$ & $80$ & $160$ & $320$\\
\hline 
\hline 0/33 & $11.11\%$ & $19.44\%$ & $36.11\%$ & $55.56\%$ & $83.33\%$ & $91.67\%$ & $100.00\%$ \\
 1/2 & $100.00\%$ & $100.00\%$ & $100.00\%$ & $100.00\%$ & $100.00\%$ & $100.00\%$ & $100.00\%$ \\
 2/4 & $50.00\%$ & $75.00\%$ & $75.00\%$  & $75.00\%$  & $75.00\%$  & $75.00\%$  & $100.00\%$ \\
 3/3 & $100.00\%$ & $100.00\%$ & $100.00\%$ & $100.00\%$ & $100.00\%$ & $100.00\%$ & $100.00\%$ \\
 4/31 & $18.18\%$ & $27.27\%$ & $42.42\%$ & $69.70\%$ & $81.82\%$ & $100.00\%$ & $100.00\%$ \\
 5/6 & $66.67\%$ & $66.67\%$ & $83.33\%$ &  $83.33\%$ & $83.33\%$ & $100.00\%$ & $100.00\%$ \\
 6/5 & $60.00\%$ & $80.00\%$ &  $80.00\%$ &   $80.00\%$ &  $100.00\%$ &  $100.00\%$ &  $100.00\%$  \\ 
 7/5 & $60.00\%$ & $80.00\%$ &  $80.00\%$ &   $80.00\%$ &  $80.00\%$ &  $100.00\%$ &  $100.00\%$  \\ 
 8/5 & $60.00\%$ & $60.00\%$ &  $80.00\%$ &   $80.00\%$ &  $80.00\%$ &  $100.00\%$ &  $100.00\%$  \\ 
 9/5 & $50.00\%$ & $50.00\%$ &  $100.00\%$ &   $100.00\%$ &  $100.00\%$ &  $100.00\%$ &  $100.00\%$  \\ 
\hline 
\end{tabular}
%}
\label{table:bin_values}
\end{table*}

\vspace{0.05in}
\noindent\textbf{Bin mapping.} 
Let  $B_j^k$ be the $j$-th bin of the $k$-th  feature  and denote the leaked  partition by 
$L(B_j^k) =\left\{ \mathbf{x}_i^k | \text{$\mathbf{x}_i^k$ is partitioned into  $B_j^k$ and $B_j^k$ is leaked} \right\}$. Here, we seek to infer the left and right bounds of $B_j^k$ from the leaked partition $L(B_j^k)$ with the help of some auxiliary data. It is easy for the adversary to simulate several samples in real-world applications and use them as auxiliary data.

We assume the adversary can obtain a few samples with full features, including the features held by the target participant $\mathcal{B}$. We call these few samples the auxiliary data which is denoted by $Aux=\left\{\mathbf{x}_i | \text{$\mathbf{x}_i$'s features are  known to the adversary} \right\}$. Given 
$Aux$ and $L(B_j^k)$, the adversary can roughly guess the left and right bounds of $B_j^k$ as $\min_{\mathbf{x}_i \in L(B_j^k) \cap Aux} \mathbf{x}_i^{k}$ and $\max_{\mathbf{x}_i \in L(B_j^k) \cap Aux} \mathbf{x}_i^{k}$.

% Let  $P^k=\left\{\mathbf{x}_i |\text{$\mathbf{x}_i$'s partial order on $k$-th dimension is leaked} \right\}$ denote the sample set with leaked information of partial orders on $k$-th dimension.  Given $L(B_j^k)$, we define $P_{j}^k=\left\{x_i | \mathbf{x}_i \in P^k \text{~and~}  \mathbf{x}_i  \text{~is partitioned into $B_j$}\right\}$ as the set of examples whose partial orders and raw features are known to the adversary. Now, the adversary infer the left and right bound of each bin $B_j$ as $\min_{\mathbf{x}_i \in P_j^k} \mathbf{x}_i^{k}$ and $\max_{\mathbf{x}_i \in P_j^k} \mathbf{x}_i^{k} $.  

We experimentally evaluate on the Credit dataset the above mapping approach to learn the left and right bounds of all the bins. We use the default segmentation to get bins on all the features of the Credit dataset.  Table~\ref{table:bin_values} shows the percentage of the successfully inferred bin values (left and right bounds of the bin) with different sizes (ranging from 5 to 320 samples) of the auxiliary dataset. As shown in the table, for a particular feature space, the number of auxiliary samples required to infer the bounds depends on the number of data bins. The more data bins used in a feature space, the more number of auxiliary samples will be required to infer the bin values up to a satisfactory accuracy.

From Table~\ref{table:bin_values}, we can see that the adversary can successfully infer the left and right bounds of each bin if it knows the whole features of a few samples.

\begin{table}[!h]
\centering
\caption{Performance of the alternative classifiers. $\#$ denotes the number of known samples used to estimate bin values. }
%\resizebox{0.45\textwidth}{!}{
\begin{tabular}{c|c|c} 
\hline \textbf{Classifier} & \textbf{$\#$samples} & \textbf{Accuracy} \\
\hline
\hline  
$C_0$ & $-$ & $81.23\%$ \\
$C_1$ & $10$ & $81.23\%$ \\ 
$C_2$ & $20$ & $81.23\%$ \\
$C_3$ & $40$ & $81.23\%$ \\
$C_4$ & $80$ & $81.23\%$ \\
$C_5$ & $160$ & $81.23\%$ \\
$C_6$ & $320$ & $81.23\%$ \\
\hline 
\end{tabular}
%}
\label{table:alternative_classifier}
\end{table}

\vspace{0.05in}
\noindent \textbf{Training alternative classifiers on leaked information.} Next, we train an alternative classifier on the Credit dataset using the leaked information. We simulate the participants in a typical VFL setting and train a distributed model using the combination of participant $\mathcal{A}$'s raw training data and participant $\mathcal{B}$'s leaked partial orders. In other words, we simply use the inferred bin values as the bin values of $\mathcal{B}$'s private data.

Table~\ref{table:alternative_classifier} compares the performance of the alternative classifier with the one trained on the raw data. $C_0$ represents the original distributed classifier  trained on the raw data, while $C_1$ to $C_6$ are six alternative classifiers trained with bin values estimated using different number of auxiliary samples. One can observe that the alternative classifiers can achieve the same classification accuracy as the original one. This result indicates that the leaked information has high potential commercial value, demonstrating the severity of privacy leakage under our reverse sum attack.
% raising practical concerns on the privacy leakage in VFL under our reverse sum attack.

\section{Possible Defenses}
\label{sec:countermeasure}
In this section, we discuss several potential defenses that may mitigate the proposed two attacks.

\vspace{0.05in}
\noindent\textbf{Robust hyper-parameter setup before training}. Our experiments in $\S$~\ref{reverse_sum_attack_exp} showed that: (\romannumeral1) if the batch size is small, the reverse multiplication attack can accurately infer all the private data of the target participant;  and (\romannumeral2) if we increase the learning rate,  the attack performance could be greatly improved. These observations indicate that one can use a large batch size and a small learning rate to mitigate the reverse multiplication attack. Of course, this comes at a cost of communication bandwidth and training efficiency. Whilst a large batch size requires a large network bandwidth for the two participants to communicate, a small learning rate could greatly increase the training time. Future work can explore new VFL mechanisms for efficient training with large batch size and small learning rate, as a defense to the reverse multiplication attack.

\vspace{0.05in}
\noindent \textbf{Hiding partial orders.}  One possible defense for SecureBoost is to hide the partial orders using secret sharing~\cite{wenjing2020arxiv}. That is, introducing an indicator vector for all the nodes to record which samples belong to them and transforming the indicator vector to a secret sharing variable to preserve privacy. In this way, the sum of the derivatives of the samples in a single node can be computed as the inner product of the indicator vector and the derivative vector. Though this method can protect the privacy of participants' partial orders, it might not be effective against our reverse sum attack. In the reverse sum attack, the adversary maliciously encodes redundant information to guess the target participant's training data, which cannot be prevented by secretly sharing the indicator vector.

\vspace{0.05in}
\noindent \textbf{Differential Privacy (DP)}.  DP~\cite{dwork2008differential} is a technique that allows the sharing of information about a dataset by describing the patterns of the classes while withholding information about the individuals. It cannot be applied to defend against the reverse sum attack since it only protect the privacy of the individual sensitive samples but cannot protect the privacy of the whole dataset (i.e., the partial orders). As for the reverse multiplication attack, DP might roughly protect the target participant's private sensitive dataset. Yet, the adversary can still infer the dataset that is added with DP noise, and use this dataset to train an equivalent model.

\section{Related Work}
\label{sec:related_work}

\noindent \textbf{VFL.}  
% the definition of privacy preserving machine learning
VFL is a type of privacy-preserving \emph{decentralized} machine learning and is tightly related to multi-party computation. Many research efforts have been devoted to improve both the efficiency and effectiveness of VFL~\cite{du2002building,vaidya2008,vaidya2002,vaidya2003,vaidya2004sdm,yu2006pakdd,aono2016codaspy,giacome2018,phong2018,mohassel2017sp,mohasse2018ccs}. They use either secure multi-party or homomorphic encryption and Yao's garbled circuits for privacy guarantees. For computation efficiency, most of the work chooses to sacrifice part of the data security and user privacy. 

In VFL, logistic regression and XGBoost are two widely used algorithms and various learning protocols have been proposed logistic regression~\cite{stephen2017corr,aono2016codaspy,chen2018logistic,kim2018secure} and XGBoost~\cite{secureboost,tian2020federboost,wenjing2020arxiv,feng2019securegbm,li2020practical} . 

Our work distinguishes itself from previous works by conducting the first study on the potential privacy risks in existing commercial VFL protocols.
 
 \vspace{0.05in}
\noindent \textbf{Privacy leakage in HFL.}   
The privacy leakage issues of HFL have attracted significant attention~\cite{melis2019sp,nasr2019sp,hitaj2017ccs,wang2019infocom,wen2020,zhaorui2019,phong2017,jonas2020,pustozerova2020information}. Existing privacy leakage research can be further categorized into three types: 1) membership~\cite{nasr2019sp,pustozerova2020information}, 2) data properties~\cite{melis2019sp}, and 2) data leakage~\cite{hitaj2017ccs,zhu2019nips,wang2019infocom,jonas2020,zhaorui2019,wen2020}. 

In membership leakage, the adversarial participant desires to learn whether a given sample is part of the other participant's training data. Nasr et al.~\cite{nasr2019sp} conducted a comprehensive analysis of the membership inference attack in FL, showing part of the training data membership is leaked when collaboratively training a model between multiple participants. Pustozerova et al.~\cite{pustozerova2020information} investigated the membership attack against sequential data.

In data properties leakage, the adversarial participant tries  to infer data properties that are independent of the main learning task, e.g., inferring whether people wearing a glass in an age prediction task.  Melis et al.~\cite{melis2019sp} showed that the gradient updates leak unintended information about participant's training data and developed passive and activate property inference attack to exploit this leakage.  

In data leakage, the adversary attempts to reconstruct the representatives of the model's training data~\cite{hitaj2017ccs,zhu2019nips,wang2019infocom,jonas2020}. 
Several works~\cite{hitaj2017ccs,wang2019infocom} have exploited the Generative Adversary Networks (GANs) to generate prototypical samples of the target participant's training data during the federated learning process. It turns out that the complete training data can even be reconstructed from the shared gradient updates. Zhu et al.~\cite{zhu2019nips} proposed an optimization algorithm that can extract both training samples and the labels from the gradient updates. Geiping et al.~\cite{jonas2020}  showed that it is relatively easier to reconstruct images at high resolution from the gradient update.

%~\cite{melis2019sp,nasr2019sp,hitaj2017ccs,wang2019infocom}, 
The above works all focus on HFL and have shown that HFL does not protect the data privacy of honest participants. Compared with HFL,  VFL poses even more challenges due to the significantly higher complexity of the learning protocol. 

\vspace{0.05in}
\noindent \textbf{Privacy leakage in VHL}. Recently,  a few works turn to study the problem of privacy leakage issue in VFL. For example,  Luo et al.~\cite{luo2020feature} proposed two specific attacks on the logistic regression and decision tree models using the model outputs. Their attacks can learn the correlations between the adversary's features and the attack target's features based on multiple predictions accumulated by the adversary. Different from \cite{luo2020feature} that learns data correlations, our two attacks aim to infer private features of the target participant.

\section{Conclusion and Discussion}
\label{sec:conclusion}
\subsection{Conclusion}
In this paper, we designed two privacy attacks against two popular vertical federated learning (VFL) protocols, i.e., reverse multiplication attack against secure logistic regression and reverse sum attack against SecureBoost. Using the two proposed attacks, we revealed the possibility of stealing private information in secured VFL frameworks. Extensive experiments on multiple benchmark datasets with two popular real-world VFL frameworks show the effectiveness of our proposed attacks and the value of the leaked information. Our results also indicate that over pursuit of efficiency may increase the privacy risks in VFL. We also discussed potential countermeasures to the two proposed attacks and their challenges. We hope our work can help the community build more advanced privacy-preserving VFL protocols.

\subsection{Limitations and Future Work}
We propose two attacks, the reverse multiplication attack and the reverse sum attack. We have evaluate the effectiveness of our attacks against two-party VFL collaboration. We do not evaluate on multi-party VFL collaboration since (\romannumeral1) the practical scenario usually involve only two enterprises.  In real world, it is almost impossible to coordinate three or more enterprises for joint modeling due the cost of computing and network bandwidth; (\romannumeral2) most of the practical VFL are designed for two-party VFL collaboration. We believe our attack can work on multi-party scenarios or can be easily extended to multi-party scenarios. Investigating the privacy issue of multi-party collaboration is a good research  direction if  multi-party VFL collaboration becomes popular in real world.

We implement the secure logistic regression and the SecureBoost protocols in \texttt{FATE}. We evaluate our attacks against the implementation built upon \texttt{FATE}.  Our attacks are supposed to be effective under any VFL frameworks, including \texttt{PySyft}, \texttt{CrypTen} and \texttt{TF Federated}. In the future, we plane to implement the secure logistic regression and SecureBoost protocols using the other VFL frameworks and evaluate our attacks against the implementations.

\bibliographystyle{plain}
\bibliography{reference}

\end{document}